\documentclass[aps,prd,twocolumn,amsmath,10pt,superscriptaddress,floatfix,nofootinbib]{revtex4-1}

\usepackage{epsfig,amssymb,amsfonts,amsmath,mathtools,bm,color,xcolor,graphicx}

\usepackage{hyperref}
\usepackage{multirow}
\hypersetup{pdftex,colorlinks=true,linkcolor=blue,citecolor=blue,menucolor=black,urlcolor=blue,filecolor=blue}

\graphicspath{{Figures/}}

\begin{document}

\title{Pentaquarks $P_c$ in a dynamical coupled-channel approach of $\gamma p \to J/\psi p$ reaction}
\author{Xu Zhang}\email{zhangxu@itp.ac.cn}
\affiliation{CAS Key Laboratory of Theoretical Physics, Institute of Theoretical Physics,\\
Chinese Academy of Sciences, Beijing 100190, China}

\begin{abstract}
We construct a dynamical coupled-channel approach to photon-induced $J/\psi$ production to search for the trace of the pentaquarks $P_c$.
In this approach, the considered open-charm intermediate states are $\Lambda_c \bar{D}^{(*)}$ and $\Sigma_c \bar{D}^{(*)}$. Scattering amplitude involving $\Lambda_c \bar{D}^{(*)}$, $\Sigma_c \bar{D}^{(*)}$ and $J/\psi p$ interactions is calculated from solving a set of coupled-channel integral equations. The coupled-channel interaction potentials are generated from one-boson-exchange.
The pentaquarks $P_c$ are dynamically generated from the coupled-channel interactions.
Moreover, the background contributions to the $P_c$ signals are assumed to be the Pomeron exchange and $\sigma$ exchange. The predicted total and differential cross sections of $J/\psi$ photoproduction are within the range of the data from the GlueX and $J/\psi$-$007$ experiments. It is found that the $\sigma$ exchange plays a crucial role for determining the very near threshold cross section. The total cross section gets enhanced due to the open-charm state rescattering in the $P_c(4312)$, $P_c(4440)$ and $P_c(4457)$ production regions, and is of order 1 nb.

\end{abstract}

\pacs{xxxxxxxxxx}

\maketitle

\section{Introduction}
The near threshold $J/\psi$ photoproduction has a long history~\cite{Gittelman:1975ix,Camerini:1975cy}. These processes may provide a window to probe the gluonic structure in the nucleon and also an effective and promising candidate process for searching of the multi-quark states. Recently, the $J/\psi$-$007$ experiment in Hall C of Jefferson Laboratory~\cite{Duran:2022xag} and 
the GlueX experiment in Hall D~\cite{GlueX:2019mkq,GlueX:2023pev} have reported new data for near threshold $J/\psi$ photoproduction.  

Assuming suppressed contributions from the intrinsic heavy quarks of proton, $J/\psi$ photoproduction is dominated by the exchange of gluons. Using the Vector Meson Dominance (VMD) model, the process $\gamma p \to J/\psi p$ is related to forward $J/\psi p \to J/\psi p$ scattering~\cite{Gell-Mann:1961jim,Kroll:1967it,Kuraev:1977fs,Balitsky:1978ic,Bauer:1977iq,Pumplin:1975fd,Barger:1975ng}. The $J/\psi p$ scattering can be expressed in terms of gluonic matrix elements using the operator product expansion~\cite{Peskin:1979va,Bhanot:1979vb,Voloshin:1978hc,Gottfried:1977gp,Appelquist:1978rt,Luke:1992tm}.
These processes provide crucial information about the trace anomaly contribution to the proton mass~\cite{Kharzeev:1995ij,Kharzeev:1998bz,Ji:1994av,Hatta:2018ina,Kou:2023zko}. Moreover, in Refs.~\cite{Luke:1992tm,Brodsky:1997gh,wu:2024xxx}, the $J/\psi$-nucleon dynamics were suggested to be dominated by the QCD van der Waals force
mediated by multi-gluon exchange. These interactions are expected to be attractive at short distance, possibly generate
a bound state between the $J/\psi$ and a light nucleus~\cite{Brodsky:1989jd,Strakovsky:2019bev,Pentchev:2020kao}.

Attempts have been made to compute near threshold $J/\psi$ photoproduction in perturbative QCD~\cite{Brodsky:2000zc,Ivanov:2004vd}.
The large mass of the heavy quark provides a hard scale, which
is much larger than the nonperturbative scale $\sim \Lambda_{\rm QCD}$. The separation of the scales justifies the application of the QCD factorization that allows one to evaluate the amplitude by convolving the perturbatively calculable hard scattering amplitude and non-perturbative quantities~\cite{Collins:2011zzd}. In Refs.~\cite{Ivanov:2004vd,Guo:2021ibg,Tong:2021ctu}, utilizing the heavy quark limit, it was shown that $J/\psi$ photoproduction can be factorized at threshold and large transferred momentum $|t|$ in terms of the gluon generalized parton distributions,
extending the factorization for the diffractive process at high energy and small $|t|$~\cite{Frankfurt:2002ka,Collins:1996fb}. This leads to lots of theoretical interests aiming to study the proton gravitational form factors in such processes~\cite{Kharzeev:2021qkd,Kharzeev:2021qkd,Sun:2021gmi,Guo:2023pqw,Guo:2023qgu,Pentchev:2024sho,Mamo:2022eui,Hatta:2023fqc}.

However, it is remarkable that, a large cross section of the open-charm state production
suggests the open-charm channel contributions may be relevant to near threshold $J/\psi$ photoproduction~\cite{Huang:2016tcr,Du:2020bqj,Skoupil:2020tge,JointPhysicsAnalysisCenter:2023qgg}. 
In addition, the hidden-charm pentaquark candidates $P_c(4312)$, $P_c(4440)$ and $P_c(4457)$ reported by LHCb Collaboration were observed from the analysis of the $J/\psi p$ mass spectrum in $\Lambda_b \to J/\psi p K^-$ decay~\cite{LHCb:2015yax,LHCb:2016ztz,LHCb:2019kea}. Closeness of the $\Sigma_c \bar{D}$, $\Sigma_c \bar{D}^*$ thresholds to these three structures suggests that $P_c(4312)$ can be assigned as an $S$-wave $\Sigma_c \bar{D}$ bound state with spin parity $\frac{1}{2}^-$,  
$P_c(4440)$ and $P_c(4457)$ as $S$-wave $\Sigma_c \bar{D}^*$ bound states with spin parity either $\frac{1}{2}^-$ and $\frac{3}{2}^-$, or $\frac{3}{2}^-$ and $\frac{1}{2}^-$, respectively~\cite{Wu:2010jy,Liu:2019tjn,Fernandez-Ramirez:2019koa,Du:2019pij,Chen:2019bip,Chen:2019asm,He:2019ify,Guo:2019kdc,Xiao:2019mvs,Wang:2019ato,Xiao:2019aya,Meng:2019ilv,Voloshin:2019aut,Wang:2019hyc,Yamaguchi:2019seo,Lin:2019qiv,Gutsche:2019mkg,Burns:2019iih,Zhu:2019iwm,Wang:2019spc,Du:2021fmf,Shen:2024nck}. A large coupling between the open-charm channels and the pentaquarks suggests the pentaquark contributions may be also relevant to near threshold $J/\psi$ photoproduction. The $J/\psi$ photoproduction has generated much interest to search for the hidden-charm pentaquarks $P_c$ both in theoretically~\cite{Wang:2015jsa,Kubarovsky:2015aaa,Karliner:2015voa,HillerBlin:2016odx,Winney:2019edt,Wang:2019krd,Wu:2019adv,Cao:2019kst,Paryev:2022wov,Strakovsky:2023kqu,Clymton:2024fbf,Duan:2024hby} and experimentally~\cite{Meziani:2016lhg,GlueX:2019mkq,Duran:2021exb,GlueX:2023pev}. 

In present work, we construct a dynamical coupled-channel approach to photon-induced $J/\psi$ production to search for the trace of the pentaquarks $P_c$.
In this approach, the considered open-charm intermediate states are $\Lambda_c \bar{D}^{(*)}$ and $\Sigma_c \bar{D}^{(*)}$. Scattering amplitude involving $\Lambda_c \bar{D}^{(*)}$, $\Sigma_c \bar{D}^{(*)}$ and $J/\psi p$ interactions is calculated from solving the Blankenbecler-Sugar (BS) equation~\cite{Blankenbecler:1965gx,Aaron:1968aoz}. 
Other alternative schemes for the dynamical coupled-channel interactions can be found in Refs.~\cite{Sato:1996gk,Lutz:2001mi,Penner:2002md,Ronchen:2012eg}. The coupled-channel interaction potentials are generated from one-boson-exchange.
In our calculation, the $\Lambda_c \bar{D}^{(*)}$ and $\Sigma_c \bar{D}^{(*)}$ interaction potentials contain $\pi$, $\eta$, $\sigma$, $\rho$ and $\omega$ exchanges. 
The $\Lambda_c \bar{D}^{(*)} \to J/\psi p$ and $\Sigma_c \bar{D}^{(*)} \to J/\psi p$ transition potentials are generated from $D$ and $D^*$ exchanges. Within this approach, the pentaquarks $P_c$ are dynamically
generated from the coupled-channel interactions,
and the contribution of the pentaquarks $P_c$ to near threshold $J/\psi$ photoproduction is calculated. In present work, the main
background contributions to the $P_c$ signals are assumed to be the Pomeron exchange~\cite{Donnachie:1984xq,Laget:1994ba,Oh:2000zi,Forshaw:1997dc} and $\sigma$ exchange.

\begin{figure*}[t]
\begin{center}
\includegraphics [scale=0.2] {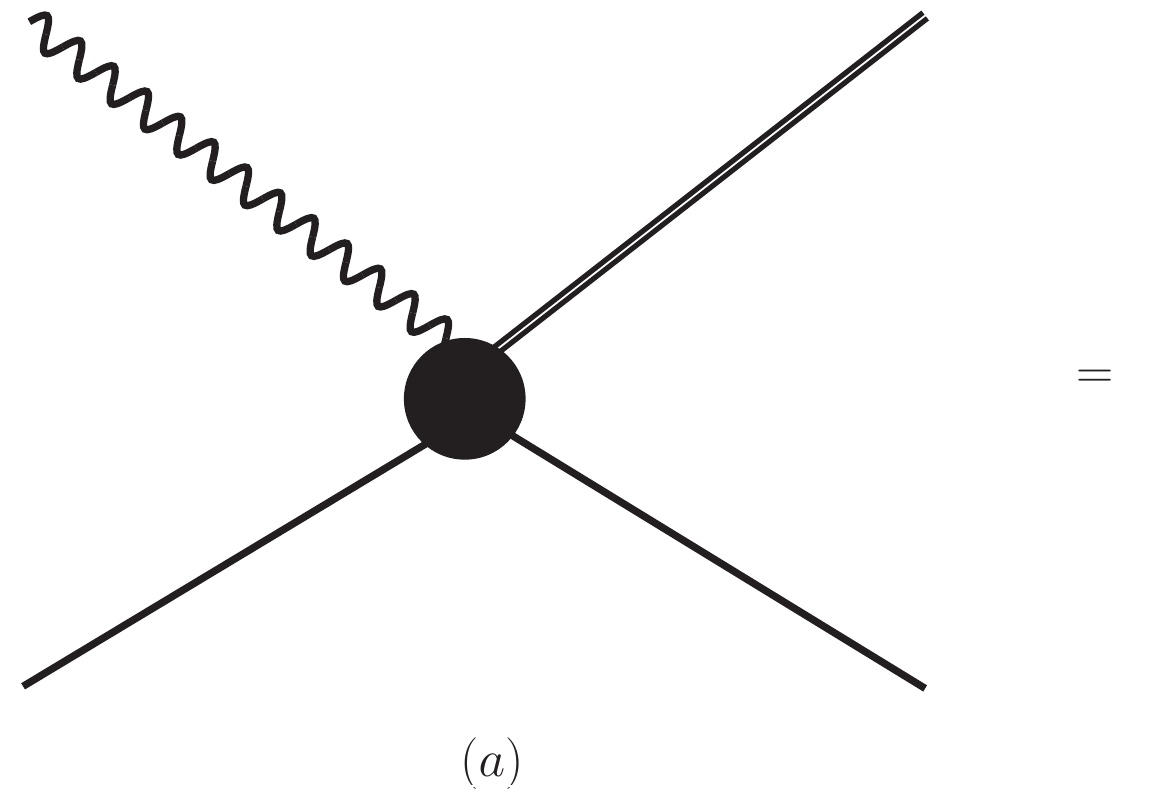}
\includegraphics [scale=0.2] {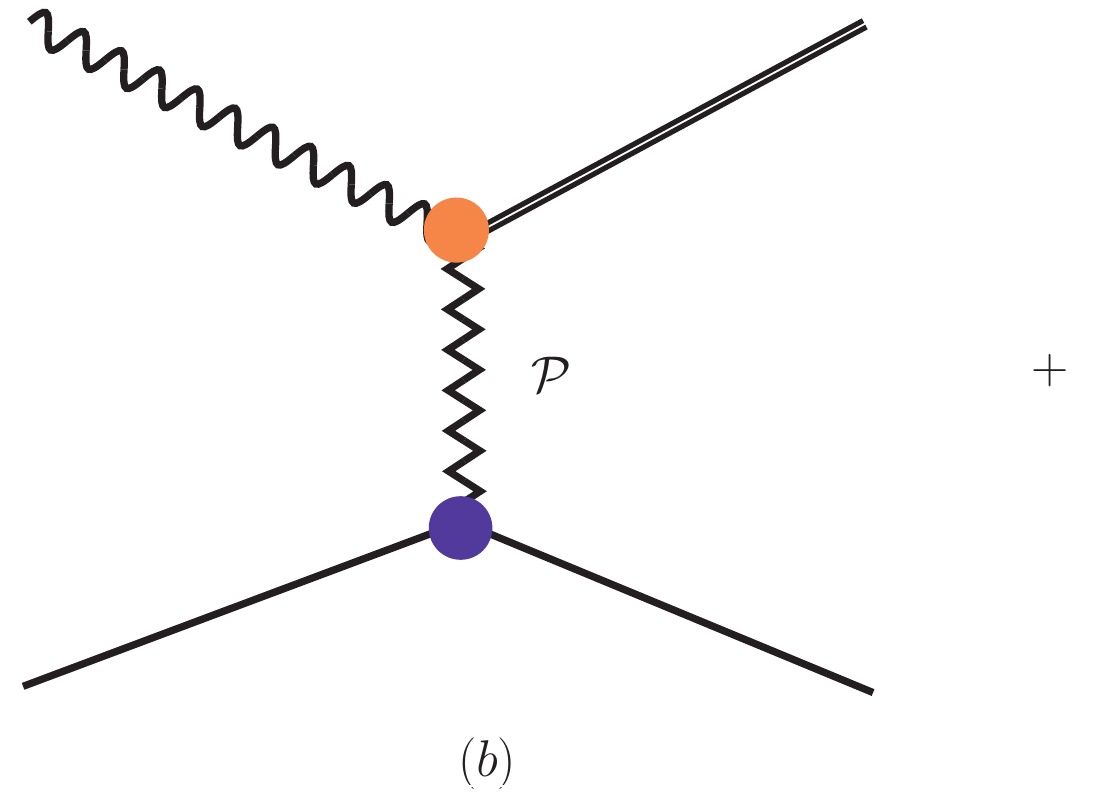}
\includegraphics [scale=0.2] {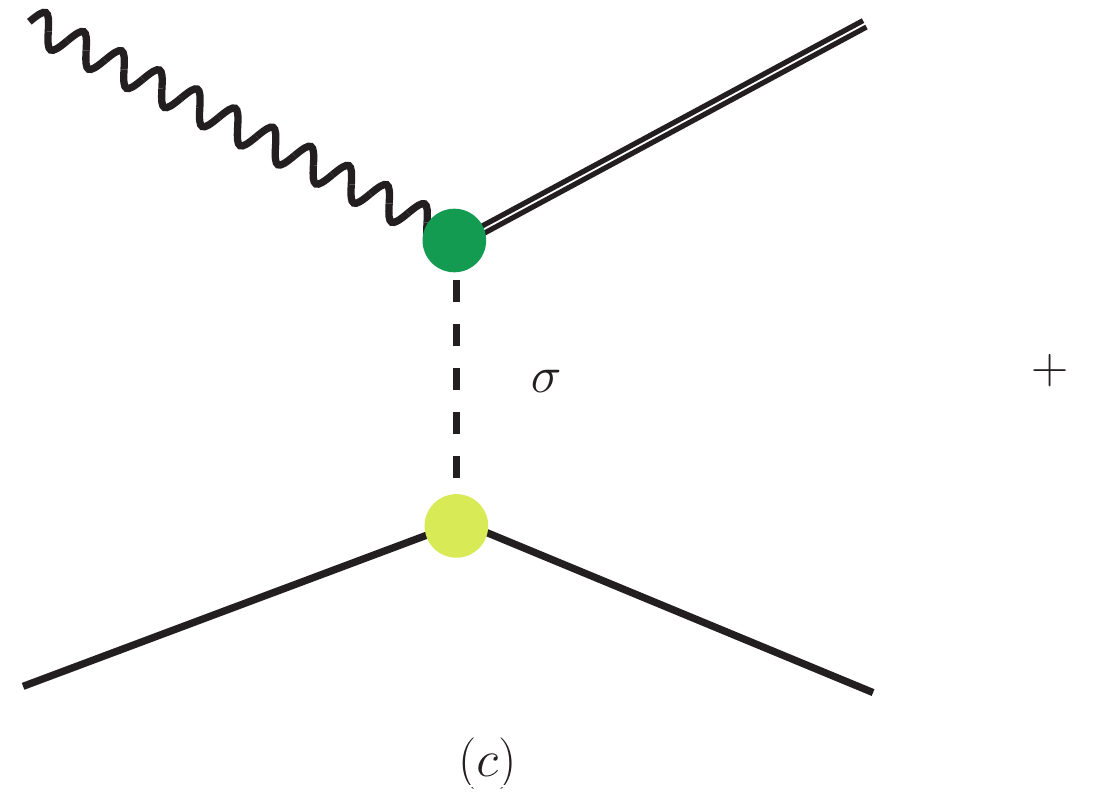}
\includegraphics [scale=0.2] {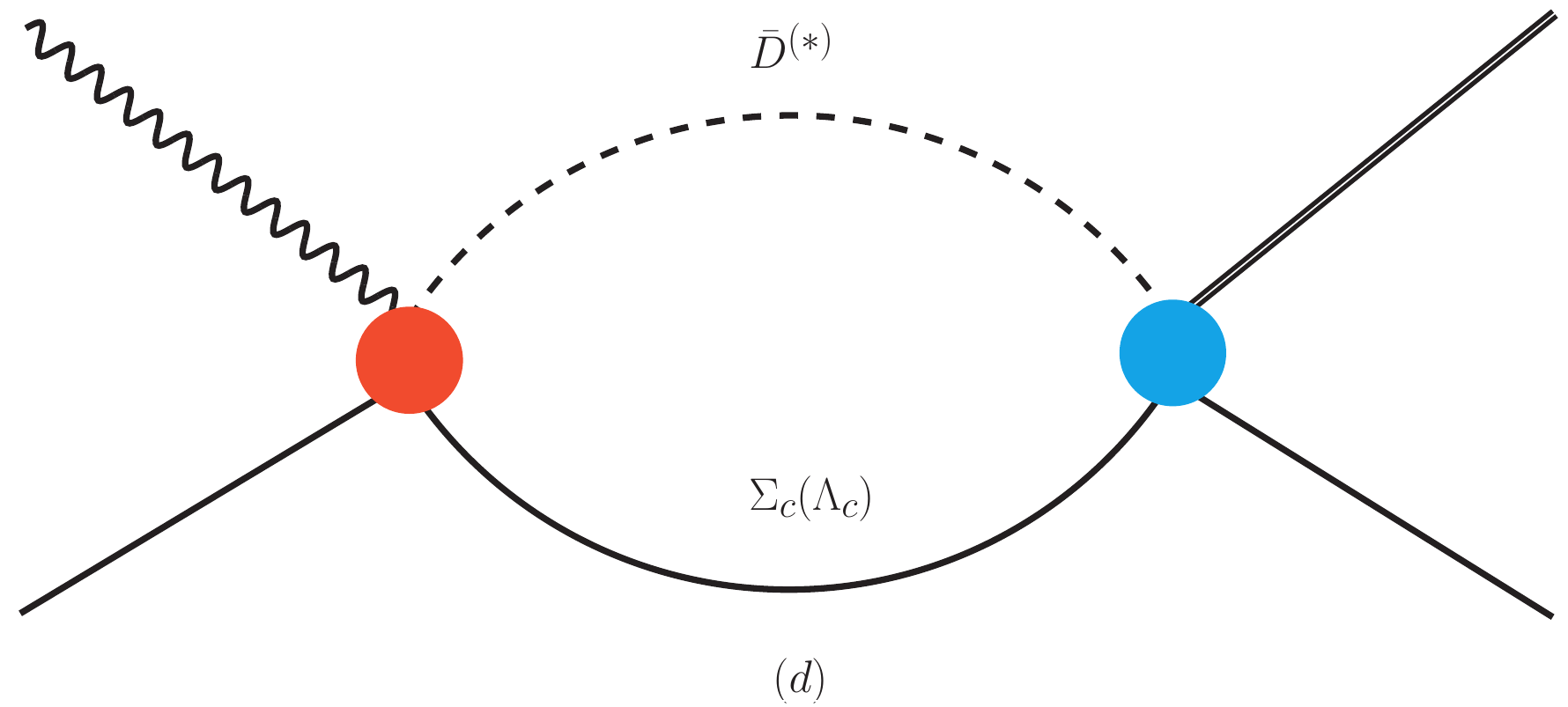}
\caption{Diagram representation of the photon-induced $J/\psi$ production.}
\label{fig:photoInduce}
\end{center}
\end{figure*}

The paper is organized as follows. In Section~\ref{sec:TheoFram}, we present the theoretical framework of this work.
In Section~\ref{sec:BackContri}, we describe the background contributions to the $P_c$ signals.
In Section~\ref{sec:IntEQ}, we present the open-charm state rescattering mechanism.
In Section~\ref{sec:lagrpoten}, the effective Lagrangians for construction of the coupled-channel $\Lambda_c \bar{D}^{(*)}$, $\Sigma_c \bar{D}^{(*)}$ and $J/\psi p$ interaction potentials are given. And using the effective
Lagrangians, the construction of the gauge invariant $\gamma p \to \Sigma_c \bar{D}^{(*)}$ and $\Lambda_c \bar{D}^{(*)}$ transition amplitudes is discussed.
In Section~\ref{sec:Results}, the numerical results are compared with the experimental data. In the last section, some conclusions are given. 

\section{Theoretical framework}
\label{sec:TheoFram}
In this section, we present the theoretical framework for the $J/\psi$ photoproduction reaction,
$\gamma(k_1) p (k_2) \to J/\psi (k_3) p (k_4)$. In the center-of-mass (c.m.) frame, the four-momenta are defined as
\begin{align}
\label{eq:MomenDefi}
&k_1=(\omega_1(k),\vec{k}\,), \quad k_2=(\sqrt{s}-\omega_1(k),-\vec{k}\,),\nonumber \\
&k_3=(\omega_3(k'),\vec{k}'\,), \quad k_4=(\sqrt{s}-\omega_3(k'),-\vec{k}'\,).  
\end{align}
Here, $\omega_i(k)=\sqrt{m_i^2+k^2}$ is the energy of a particle with mass $m_i$, and $\sqrt{s}$ denotes the invariant mass of the system.

The differential cross section can be written as 
\begin{align}
\frac{d\sigma}{dt}=\frac{1}{64\pi s}\frac{1}{k^2} \frac{1}{4}\sum_{\rm all\, spins} | \langle \vec{k}', \lambda_3\lambda_4|\mathcal{M}| \vec{k}, \lambda_1\lambda_2\rangle|^2,
\end{align} 
where $\lambda_1$ and $\lambda_3$ denote the helicities of the photon and $J/\psi$, respectively, $\lambda_2$ and $\lambda_4$ are the helicities of the initial and final protons, respectively, and $k(k')$ is the magnitude of the three-momenta $\vec{k}(\vec{k}')$.

In present work, as shown in Fig.~\ref{fig:photoInduce}, the full amplitude $\mathcal{M}$ consists of Pomeron exchange, $\sigma$ exchange and the open-charm intermediate state rescattering. Then the amplitude $\mathcal{M}$ can be written as 
\begin{align}
\mathcal{M}=\mathcal{M}^{\rm Pom}+\mathcal{M}^{\sigma}+\mathcal{M}^{\rm ocs}.
\end{align} 
In the following sections, each amplitude will be presented.

\section{Background contributions}
\label{sec:BackContri}
Recently, the Pomeron-potential model was proposed to study the $J/\psi$ photoproduction off the nucleon target from the threshold to very high energy in Refs.~\cite{Sakinah:2024cza,Lee:2022ymp}. 
The reaction model in Ref.~\cite{Tang:2024pky} exposes the $c\bar c$ component via the $\gamma \to c\bar c +{\mathcal{P}} \to J/\psi$ loop, and couples these loop constituents to the proton via the Pomeron exchange. In present work, the main
background included to the $P_c$ signals is assumed to be the Pomeron exchange and $\sigma$ exchange.

\subsection{Pomeron exchange}
Following Refs.~\cite{Donnachie:1984xq,Laget:1994ba,Oh:2000zi}, the $J/\psi$ photoproduction amplitude of the Pomeron exchange mechanism can be written as 
\begin{align}
\mathcal{M}^{\rm Pom} = \bar{u}(k_4,\lambda_4) \mathcal{T}^{\mu\nu} {u}(k_2,\lambda_2) \epsilon^*_{\mu}(k_3,\lambda_3) \epsilon_{\nu}(k_1,\lambda_1), 
\end{align} 
where $\epsilon_{\nu}(k_1,\lambda_1)$ and $\epsilon_{\mu}^*(k_3,\lambda_3)$ are the photon and $J/\psi$ polarization vectors, respectively. And ${u}(k_2,\lambda_2)$ and $\bar{u}(k_4,\lambda_4)$ are the incoming and outgoing nucleon spinors, respectively, with the normalization $\bar{u}(k,\lambda')u(k,\lambda)=2m_N\delta_{\lambda\lambda'}$, and $m_N$ is the
nucleon mass. $\mathcal{T}^{\mu\nu}$ can be written as 
\begin{align}
\mathcal{T}^{\mu\nu} =  i \frac{12 em_{J/\psi}^2}{f_{J/\psi}} G_\mathbb{P}(s,t) \beta_{c}F_V(t) \beta_{u/d} F_1(t)
\left( k_1 \!\!\!\!\!\!/  \,\,\, g^{\mu\nu} -k_1^{\mu}\gamma^{\nu}\right),
\label{eq:MP}
\end{align}
with $t = (k_3-k_1)^2 = (k_2-k_4)^2$. 
Here, $e$ is the unit electric charge, $m_{J/\psi}$ and $f_{J/\psi}$ are the mass and decay constant of $J/\psi$, $\beta_c$ is the coupling of the Pomeron with the charm quark $c$ (or the anti-charm quark $\bar{c}$\,) in the $J/\psi$, and $\beta_{u/d}$ is the coupling of the Pomeron with the light quarks $u$ (or $d$\,) in the nucleon. 
The value of $f_{J/\psi}=11.2$ is determined through the decay width of $\Gamma(J/\psi \to e^+ e^-) = 4\pi m_{J/\psi} \alpha_{\rm em} / (3 f_{J/\psi}^2)$~\cite{ParticleDataGroup:2024cfk}, with $\alpha_{\rm em} = e^2/(4\pi)$.

The Pomeron-vector-meson vertex is dressed by the form factor,
\begin{align}
F_V(t)=\frac{1}{m_{J/\psi}^2-t} \left( \frac{2\mu_0^2}{2\mu_0^2 + m_{J/\psi}^2 - t} \right).
\end{align}
The form factor for the Pomeron-nucleon vertex can be written as
\begin{align}
F_1(t) = \frac{4m_N^2 - 2.8 t}{(4m_N^2 - t)(1-t/0.71)^2}.
\end{align}
The propagator $G_{\mathbb{P}}$ of the Pomeron follows the Regge phenomenology and can be written as 
\begin{align}
G_\mathbb{P} = \left(\frac{s}{s_0}\right)^{\alpha_P^{}(t)-1} \exp\left\{ - \frac{i\pi}{2} \left[ \alpha_P^{} (t)-1 \right] \right\} ,
\end{align}
where $ \alpha_P^{} (t) = \alpha_0^{} + \alpha'_P t$.
In the present work, the values of the parameters used in the calculation are taken from Refs.~\cite{Sakinah:2024cza,Lee:2022ymp}, 
\begin{eqnarray}
&& \mu_0^{} = 1.1~{\rm{GeV}}^2, \quad \beta_{c} = 0.32~{\rm{GeV}}^{-1}, 
\nonumber \\ &&
\alpha'_P = 1/s_0^{} = 0.25~{\rm{GeV}}^{-2}, \quad \alpha_0^{} = 1.25.
\end{eqnarray}
These parameters are determined by fitting the data of $\gamma p \to J/\psi p$ total cross section from threshold up to $300$~GeV~\cite{ZEUS:1995kab,H1:1996gwv,Gittelman:1975ix,Camerini:1975cy}.

\subsection{$\sigma$ meson exchange}

Assuming that $J/\psi\to \pi^0\pi^0 \gamma$ decay is dominated by the $\sigma$,
a large branching ratio ${\rm Br}(J/\psi\to \pi^0\pi^0 \gamma)$~\cite{ParticleDataGroup:2024cfk} suggests that the $\sigma$ exchange could have contribution to $J/\psi$ photoproduction. The coupling constant $g_{J/\psi \sigma \gamma}$ is related to the internal content of the $\sigma$.
There are many
models on the structure of the $\sigma$ meson, the nature of the $\sigma$ has not been well defined~\cite{Pelaez:2015qba}.
In present work, the coupling $g_{J/\psi \sigma \gamma}$ is
determined from the analysis of the $J/\psi$ radiative decay in Ref.~\cite{Sarantsev:2021ein}.

The electromagnetic interaction Lagrangian for the $\sigma$ exchange can be written as 
\begin{align}
\mathcal{L}_{J/\psi\sigma\gamma}=\frac{eg_{J/\psi\sigma\gamma}}{m_{J/\psi}}F^{\mu\nu}J/\psi_{\mu\nu}\sigma,
\end{align}
where $F^{\mu\nu}=\partial^{\mu}A^{\nu}-\partial^{\nu}A^{\mu}$, $J/\psi_{\mu\nu}=\partial_{\mu}J/\psi_{\nu}-\partial_{\nu}J/\psi_{\mu}$.

The coupling constant is determined by the radiative decay width of $J/\psi\to \sigma \gamma$,
\begin{align}
\Gamma_{J/\psi\to \sigma \gamma}=\frac{e^2}{3\pi}\frac{q_{\gamma}^3}{m_{J/\psi}^2}g_{J/\psi \sigma \gamma}^2,
\end{align}
where $q_{\gamma}=(m_{J/\psi}^2-m_{\sigma}^2)/(2m_{J/\psi})$.
Using the radiative branch ratio ${\rm Br}({J/\psi\to \sigma \gamma})=1.14\times 10^{-3}$ given in Ref.~\cite{Sarantsev:2021ein}, we can get $g_{J/\psi \sigma \gamma}=5.5\times 10^{-3}$. In present work, we have taken $m_{\sigma}=0.6$~GeV.

The strong interaction Lagrangian is 
\begin{align}
\mathcal{L}_{NN\sigma}=g_{NN\sigma}NN\sigma.
\end{align}
Following Ref.~\cite{Ronchen:2012eg}, the coupling constant used in present work is taken to be
$g_{NN\sigma}=13.85$.

Then the amplitude for $\sigma$ exchange can be written as 
\begin{align}
\mathcal{M}^{\sigma}=\frac{e}{m_{J/\psi}}\frac{2g_{J/\psi \sigma \gamma}g_{NN\sigma}}{t-m_{\sigma}^2}(k_1\cdot k_3 g^{\mu\nu}-k_1^{\mu}k_3^{\nu})\nonumber  \\
\times \bar{u}(k_4,\lambda_4) {u}(k_2,\lambda_2) \epsilon^*_{\mu}(k_3,\lambda_3)\epsilon_{\nu}(k_1,\lambda_1).
\end{align}

\section{Open-charm state rescattering}
\label{sec:IntEQ}
In this Section, we construct a dynamical coupled-channel approach to photon-induced $J/\psi$ production through the open-charm intermediate state rescattering. 
In this approach, $\gamma p$ produces the open-charm intermediate pairs first which then rescatter into the final $J/\psi p$. The considered open-charm intermediate states are $\Lambda_c \bar{D}^{(*)}$ and $\Sigma_c \bar{D}^{(*)}$. Scattering amplitude involving $\Lambda_c \bar{D}^{(*)}$, $\Sigma_c \bar{D}^{(*)}$ and $J/\psi p$ interactions is calculated from solving a set of coupled-channel integral equations. The coupled-channel interaction potentials are generated from one-boson-exchange. 
The pentaquarks $P_c$ are dynamically generated from the coupled-channel interactions.
In Refs.~\cite{Du:2019pij,Du:2021fmf}, the inclusion of more channels has been considered, and the coupled-channel potentials were constructed by one-pion exchange as well as short-range operators constrained by heavy quark spin symmetry.

\subsection{Dynamical coupled-channel equation}
The scattering amplitudes are obtained by iterating these potentials by using a coupled-channel BS equation~\cite{Blankenbecler:1965gx,Aaron:1968aoz}. The partial-wave decomposed, $\Lambda_c \bar{D}^{(*)}$, $\Sigma_c \bar{D}^{(*)}$ and $J/\psi p$ coupled-channel scattering amplitude $T$ can be written as 
\begin{align}
\label{eq:bse}
\nonumber {T}_{ij}(s,p',p)=&{V}_{ij}(s, p', p)+\sum_{i'}\int_C \frac{k^2dk}{(2\pi)^3}{V}_{ii'}(s, p', k) \\ 
&\times G_{i'}(s,k){T}_{i'j}(s,k,p),
\end{align}
in each matrix element $V_{ij}(s,p',p)$, the index $i(j)=1,2,3,4,5$ labels the particle channel ($J/\psi p=1$, $\Lambda_c \bar{D}=2$, $\Lambda_c \bar{D}^*=3$, $\Sigma_c \bar{D}=4$, $\Sigma_c \bar{D}^*=5$). The same structure holds for $T_{ij}(s,p',p)$. The incoming and outgoing momenta are denoted as $p$ and $p'$, respectively, and $s$ denotes the squared invariant mass of the two-body system. In this work, the partial-wave interaction potentials $V_{ij}(s,p',p)$ is in the $JLS$ basis, where the indices $J$, $L$ and $S$ have been suppressed. Here, $J$, $L$ and $S$ denote the total angular momentum, orbital angular momentum and total spin, respectively.

The two-body propagator for channel $i$ is denoted as $G_i(s,k)$ and can be written as 
\begin{align}
\label{eq:propaga}
G_i(s,k)=\frac{\omega_1(k)+\omega_2(k)}{2\omega_1(k)\omega_2(k)}\frac{1}{s-(\omega_1(k)+\omega_2(k))^2+i\epsilon},
\end{align}
where $\omega_1(k)$ and $\omega_2(k)$ are the on-mass shell energies for particles 1 and 2, respectively. The particles $1$ and $2$ in each channel denote meson and baryon, respectively. On shell momentum $k_0$ in channel $i$ is
\begin{align}
k_0=\frac{\sqrt{[s-(m_1+m_2)^2][s-(m_1-m_2)^2]}}{2\sqrt{s}}.
\end{align}

To obtain the partial wave interaction potentials in the $JLS$ basis, we use the method given in Refs.~\cite{Jacob:1959at,Chung:1971ri}. Firstly, in the helicity basis the relevant partial wave is extracted. We choose the incident $\vec{p}$ along the $z$-axis and outgoing $\vec{p}\,'$ to be in the $x,z$ plane. In the c.m. frame, the four-momenta have the same definitions as given in Eq.~\eqref{eq:MomenDefi}.   $k_1$ ($k_3$) and $k_2$ ($k_4$) are the four-momenta of the incoming (outgoing) meson and baryon, respectively.
The plane-wave potential is related to the partial-wave potential as
\begin{eqnarray}
\label{eq:patexp}
V_{\lambda'\lambda}(s,\vec{p}\,',\vec{p}\,) =\frac{1}{4\pi}\sum_{J} (2J+1) d_{\lambda\lambda'}^J({\rm cos}\theta)V_{\lambda'\lambda}^J(s,p',p),\nonumber  \\
\end{eqnarray}
with $\lambda=\lambda_1-\lambda_2$ and $\lambda'=\lambda_3-\lambda_4$, and $\theta$ is the angle between the outgoing momenta $\vec{p}\,'$ and incoming momenta $\vec{p}$. $\lambda_1$ ($\lambda_3$) and $\lambda_2$ ($\lambda_4$) denote the helicities of the incoming (outgoing) meson and baryon, respectively.

Using the orthogonality relation for the Wigner $d_{\lambda\lambda'}^J({\rm cos}\theta)$ functions, 
\begin{eqnarray}
\int_{-1}^{+1}d_{\lambda\lambda'}^J({\rm cos}\theta) d_{\lambda\lambda'}^{J'}({\rm cos}\theta)  \,d {\rm cos }\theta= \frac{2}{2J+1}\delta_{JJ'},
\end{eqnarray}
we can invert the above expression and obtain
\begin{eqnarray}
\label{eq:patwa}
V_{\lambda'\lambda}^J(s,p',p)=2\pi\int_{-1}^{+1}d_{\lambda\lambda'}^J({\rm cos}\theta)V_{\lambda'\lambda}(s,\vec{p}\,',\vec{p}\,) \,d \rm{cos}\theta.\nonumber \\
\end{eqnarray}

Then, the transition from the helicity to the $JLS$ representation is given by
\begin{eqnarray}
 V^{J}_{L'L}(s,p',p)=\sum_{\lambda'\lambda}\langle JL'S |J\lambda'\rangle V_{\lambda'\lambda}^J(s,p',p) \langle J \lambda|JLS\rangle,\nonumber \\
\end{eqnarray}
where 
\begin{eqnarray}
\langle JLS |J\lambda\rangle 
 =\sqrt{\frac{2L+1}{2J+1}} \langle L0S\lambda|J\lambda\rangle \langle s_1 \lambda_1 s_2 -\lambda_2 | S \lambda\rangle.
\end{eqnarray}
Here, $s_1$ and $s_2$ denote the particle spins. The symbols on the right-hand side are standard $SU(2)$ Clebsch-Gordan coefficients. In present work, only $S$-wave coupled-channel interaction is considered.

\subsection{Analytic continuation}

As a function of $\sqrt{s}$, the amplitude $T$ in Eq.~\eqref{eq:bse} has a branch point at $\sqrt{s}=m_1+m_2$ of channel $i$. We must therefore consider the complex $\sqrt{s}$-plane as a Riemann surface of multisheets, and different Riemann sheets are connected by a branch cut. At each threshold there is a branch point, at which a cut begins is fixed. For channel $i$, in the complex $k$-plane,
the physical and unphysical Riemann sheets correspond
to $0\leq{\rm arg}k_0<\pi$ and $\pi\leq{\rm arg}k_0<2\pi$, respectively. The amplitude $T$ on the unphysical Riemann sheet
can be obtained by analytically calculating the residues of the poles of the integrand as to be discussed. 

\begin{figure}[tbhp]
\begin{center}
\includegraphics [scale=0.32] {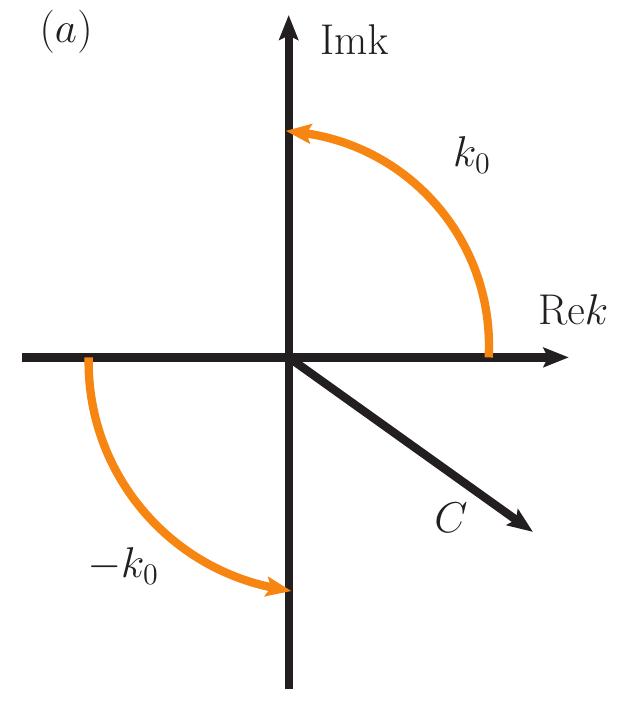}
\includegraphics [scale=0.32] {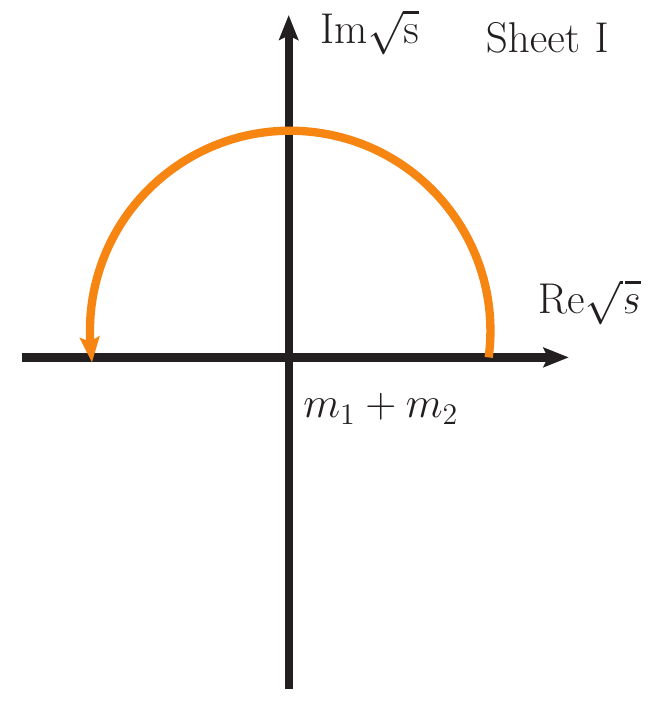}
\includegraphics [scale=0.32] {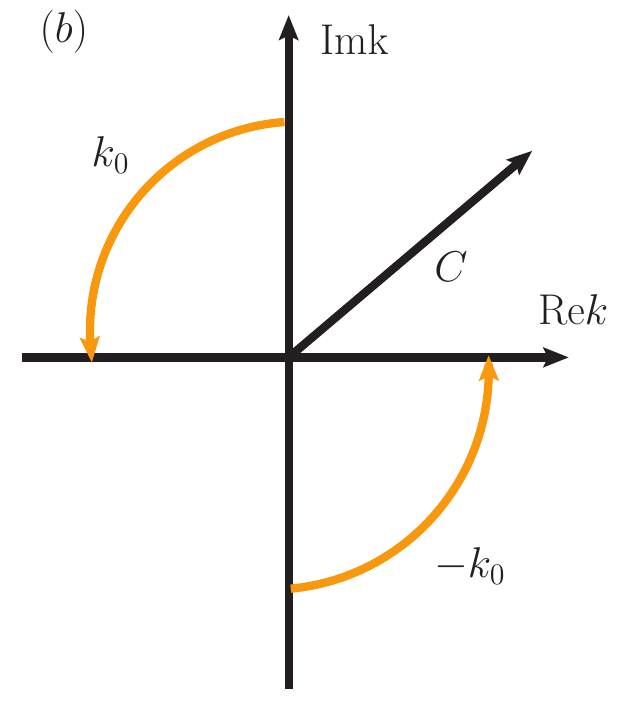}
\includegraphics [scale=0.32] {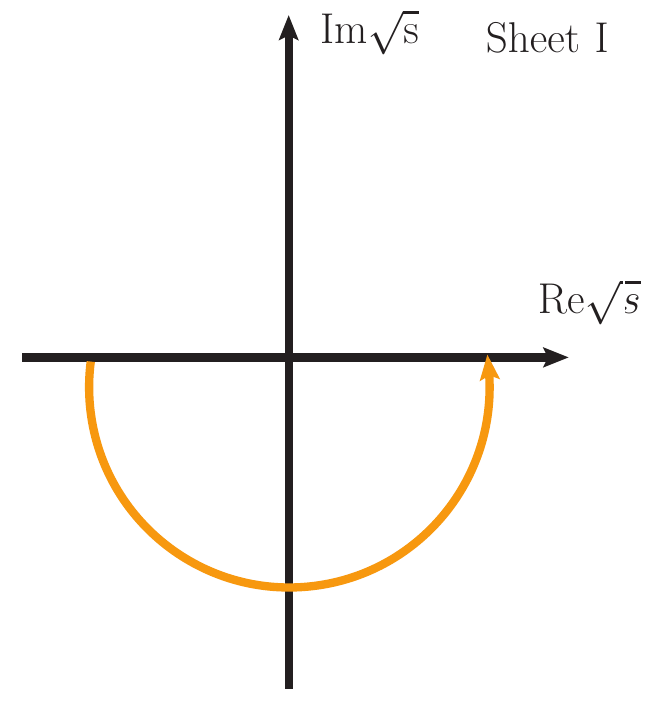}
\includegraphics [scale=0.32] {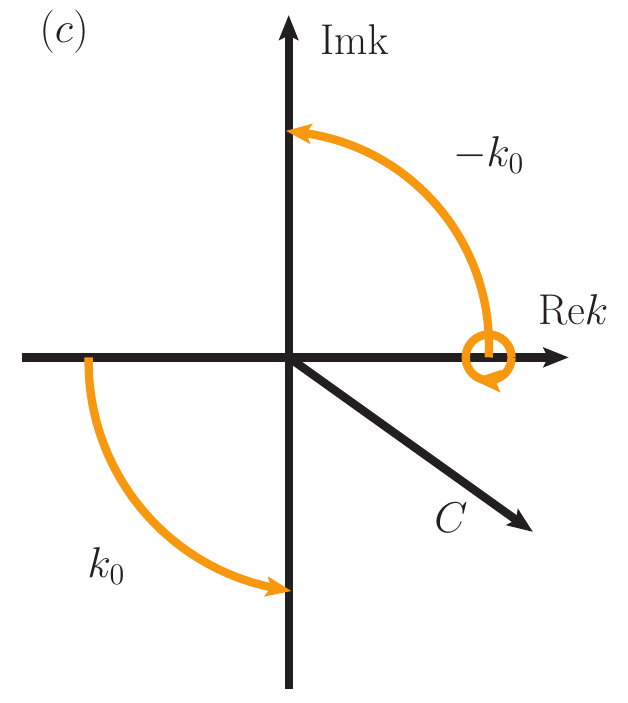}
\includegraphics [scale=0.32] {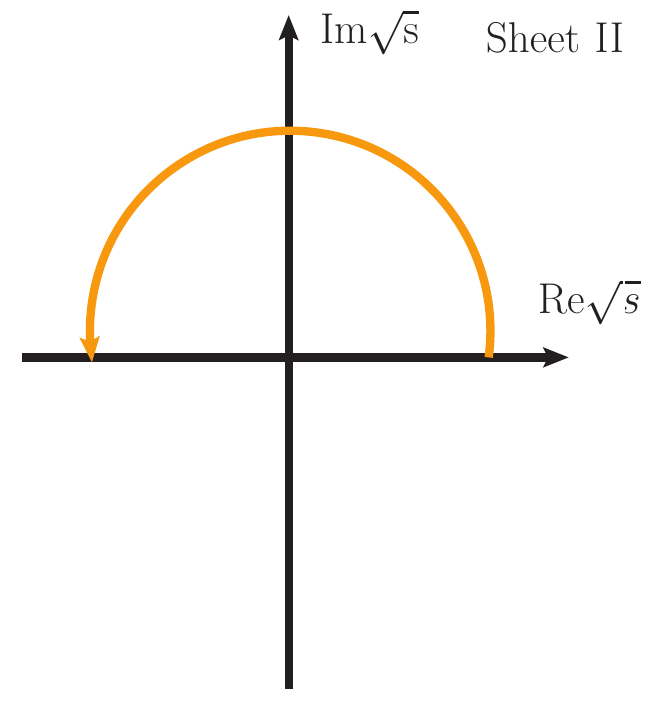}
\includegraphics [scale=0.32] {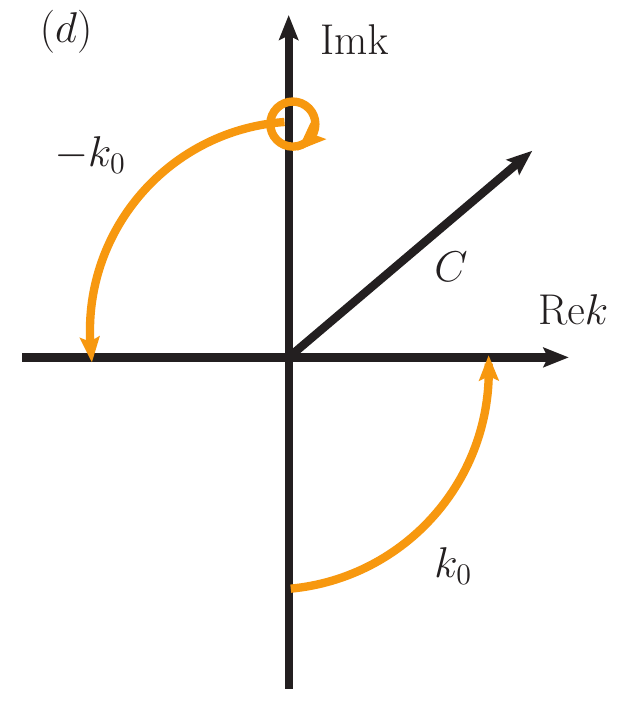}
\includegraphics [scale=0.32] {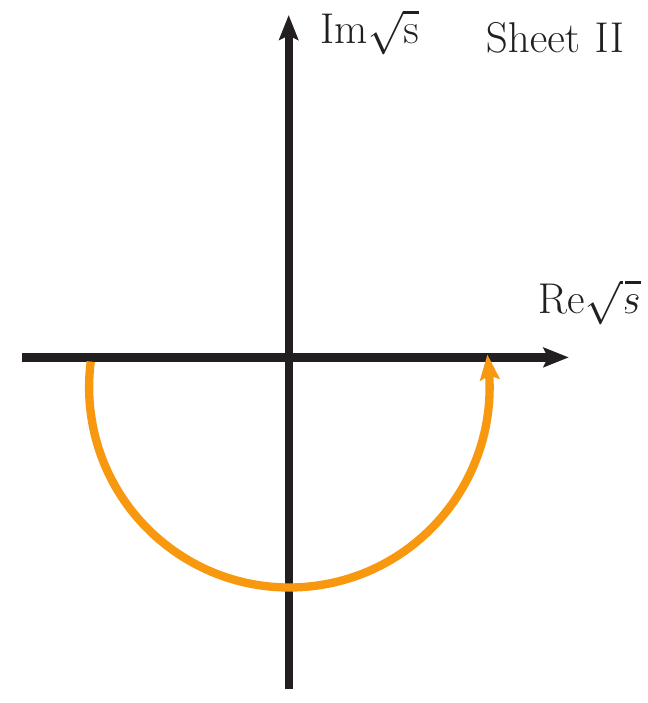}
\caption{Moving of the poles $\pm k_0$ in the complex $k$-plane corresponding to the moving of the energy $\sqrt{s}$ in the complex $\sqrt{s}$-plane. For each energy region, the integral contour $C$ is used in the calculation. Here, sheets I and II denote the physical and unphysical Riemann sheets, respectively.}
\label{fig:ContSheet}
\end{center}
\end{figure}

The physical amplitude is constrained by the $+i\epsilon$ in Eq.~\eqref{eq:propaga}. As shown in Fig.~\ref{fig:ContSheet}(a), choosing the integral contour passing the singularity at $k_0$ on the right, 
yields the amplitude on the physical Riemann sheet. As $\sqrt{s}$ moves in a counterclockwise semicircle in the complex $\sqrt{s}$-plane, the poles at $\pm k_0$ also move along counterclockwise arcs in the complex $k$-plane as indicated by the yellow curves. At the end point of this path, the poles in the complex $k$-plane lie on the imaginary axis, so that the integral contour can be deformed to that shown in Fig.~\ref{fig:ContSheet}(b). As $\sqrt{s}$ is continuously moving from the end point of the path of Fig.~\ref{fig:ContSheet}(b) onto the positions of Fig.~\ref{fig:ContSheet}(c), the amplitude $T$ is analytically continued into the unphysical Riemann sheet. The integral contour can be deformed back to the lower $k$-plane to ensure that the pole at $-k_0$ does not get too close to the integration contour. In this process, we must pick up a clockwise residue of the pole at $-k_0$. When the poles in the $k$-plane reach imaginary axis, the integral contour can be deformed to that shown in Fig.~\ref{fig:ContSheet}(d). Some detailed discussions can be found in Ref.~\cite{Pearce:1988rk}.

Then, the amplitude $T$ in the unphysical Riemann sheet for channel $i'$ is given as
\begin{align}
\label{eq:bsess}
\nonumber {T}_{ij}(s,p',p)=&{V}_{ij}(s, p', p)+\delta G + 
\sum_{i'}\int_C \frac{k^2dk}{(2\pi)^3}{V}_{ii'}(s, p', k) \\ 
&\times G_{i'}(s,k){T}_{i'j}(s,k,p),
\end{align}
where 
\begin{align}
\delta G = 2\pi i\frac{-k_0}{4(2\pi)^3\sqrt{s}} {V}_{ii'}(s, p', -k_0){T}_{i'j}(s,-k_0,p). 
\end{align}
The integral contour $C$ in Eq.~\eqref{eq:bse} and Eq.~\eqref{eq:bsess} is chosen as shown in  Fig.~\ref{fig:ContSheet}(a) for ${\rm Im}\sqrt{s} \geq 0$ and 
as in Fig.~\ref{fig:ContSheet}(b) for ${\rm Im}\sqrt{s} < 0$.

In present work, the BS equation is solved numerically by replacing the integrals by sums using Gaussian quadratures and inverting the resulting matrix equation~\cite{Haftel:1970zz}. In addition, singularities may emerge as the $\pi$ exchange potential $V(s,\vec{p}\,',\vec{p}\,)$ is projected to partial wave, e.g., for $\Sigma_c \bar{D} \to \Sigma_c \bar{D}^*$ potential.
The contour deformation has been employed as a tool to deal with these singularities~\cite{Doring:2009yv,Sadasivan:2021emk,Dawid:2023jrj,Zhang:2024dth,Feng:2024wyg,Sakthivasan:2024uwd}. 
The solution method requires the knowledge of the singularity positions in $V(s,\vec{p}\,',\vec{p}\,)$. In the multi-channel case, the solution for solving the BS equation will become complicated. In present work, we use the static approximation, that is by the replacement
\begin{align}
\frac{1}{q_t^2-m_{\pi}^2}\to \frac{1}{-\vec{q}^{\,2}_t-m_{\pi}^2},  
\end{align}
when the singularities appear in the potential $V(s,\vec{p}\,',\vec{p}\,)$. Here, the $q_t$ is the four-momenta for the exchanged pion as will be given in Eq.~\eqref{eq:Tranmome}, $m_{\pi}$ is pion mass.

\subsection{Photoproduction amplitude}
\label{sec:hadronresctter}
The partial-wave amplitude $\mathcal{M}^{\rm ocs}$ for $J/\psi$ photoproduction via the open charm $\Lambda_c \bar{D}^{(*)}$, $\Sigma_c \bar{D}^{(*)}$ rescattering can be written as
\begin{align}
\label{eq:produ}
\mathcal{M}^{\rm ocs}=\sum_{i}\frac{1}{4\pi}\int_0^{\infty}\frac{k^2dk}{(2\pi)^3}{T}_{1,i}(s, p', k)G_i(s,k){V}_{i,\gamma p}(s,k,p).
\end{align}  
The coupled-channel interaction amplitude is given in Eq.~\eqref{eq:bse}.
The open-charm photoproduction amplitude ${V}_{i,\gamma p}(s, p', p)$ will be discussed in next Section. The plane-wave potential is related to the partial-wave potential through the Eq.~\eqref{eq:patexp}. In present work, only $S$-wave open-charm state rescattering is considered.

\section{Lagrangians and potentials}
\label{sec:lagrpoten}
In this Section, the effective Lagrangians for construction of the coupled-channel $\Lambda_c \bar{D}^{(*)}$, $\Sigma_c \bar{D}^{(*)}$ and $J/\psi p$ interaction potentials are given. And using the effective
Lagrangians, the construction of the gauge invariant $\gamma p \to \Sigma_c \bar{D}^{(*)}$ and $\gamma p \to \Lambda_c \bar{D}^{(*)}$ transition potentials is discussed.

\subsection{Effective Lagrangians}
The effective Lagrangians between pseudo-Goldstone bosons and the mesons containing a heavy quark can be 
constructed by imposing invariance under both heavy quark spin-flavor transformation and chiral transformation~\cite{Burdman:1992gh,Wise:1992hn,Yan:1992gz,Falk:1992cx,Casalbuoni:1996pg}. The light vector mesons nonet can be introduced by using the hidden
gauge symmetry approach~\cite{Casalbuoni:1992dx,Casalbuoni:1992gi,Casalbuoni:1996pg}. The Lagrangians containing these particles can be written as

\begin{eqnarray}
\nonumber{\cal L}_{ DD^{*}P}&=&g_{ DD^{*}P}({
D}_b { D}^{*\mu\dagger}_a+{ D}^{*\mu}_{b}{\
D}^{\dagger}_a) (\partial_{\mu}{\cal M})_{ba} \\&&+g_{\bar{D}\,\bar{D}^{*}P}(\,\bar{
D}^{*\mu\dagger}_a\bar{
D}_b+\bar{
D}^{\dagger}_a\bar{
D}^{*\mu}_b)(\partial_{\mu}{\cal M})_{ab},\\
\nonumber{\cal L}_{D^{*}D^{*}P}&=&g_{ D^*D^{*}P}\epsilon_{\beta\mu\alpha\nu} {D}^{*\mu}_b {\stackrel{\leftrightarrow \beta}{\partial}}{D}^{*\nu\dagger}_a
(\partial^{\alpha}{\cal M})_{ba}  \\
&&- g_{ \bar{D}^*\bar{D}^{*}P}\epsilon_{\beta\mu\alpha\nu} 
\bar{D}^{*\mu\dagger}_a {\stackrel{\leftrightarrow\beta}{\partial}} \bar{D}^{*\nu}_b
(\partial^{\alpha}{\cal M})_{ab},\\
\nonumber{\cal L}_{{\rm DD}V}&=&  \\
&& \!\!\!\!\!\!\!\!\!\!\!\!\!\!\!\!\!\!\!\!\!\!\!\!ig_{{ DD}V}({
D}_b\stackrel{\leftrightarrow}{\partial_{\mu}}{
D}^{\dagger}_a)V^{\mu}_{ba}+ig_{\bar{ D}\,\bar{
D}V}(\bar{
D}_b\stackrel{\leftrightarrow}{\partial_{\mu}}\bar{
D}^{\dagger}_a)V^{\mu}_{ab},\\
\nonumber {\cal L}_{{D^{*}D^{*}}V}&=& ig_{{ D^{*}D^{*}}V}({
D}^{*}_{b\nu}\stackrel{\leftrightarrow}{\partial_{\mu}}{
D}^{*\nu\dagger}_a)V^{\mu}_{ba} \\
\nonumber&&\!\!\!\!\!\!\!\!\!\!\!\!\!\!\!\!\!\!\!\!\!\!\!\!\!\!+ig'_{{ D^{*}D^{*}}V}({
D}^{*\mu}_{b}{ D}^{*\nu\dagger}_{a}-{ D}^{*\mu\dagger}_{a}{
D}^{*\nu}_{b})(\partial_{\mu}V_{\nu}-\partial_{\nu}V_{\mu})_{ba}\\
\nonumber&&\!\!\!\!\!\!\!\!\!\!\!\!\!\!\!\!\!\!\!\!\!\!\!\!\!\!+ ig_{\bar{D}^{*}\bar{
D}^{*}V}(\bar{
D}^{*}_{b\nu}\stackrel{\leftrightarrow}{\partial_{\mu}}\bar{ D}^{*\nu\dagger}_{a})V^{\mu}_{ab}\\
&&\!\!\!\!\!\!\!\!\!\!\!\!\!\!\!\!\!\!\!\!\!\!\!\!\!\!+ig'_{\bar{D}^{*}\bar{D}^{*}V}(\bar{D}^{*\mu}_b\bar{D}^{*\nu\dagger}_a-\bar{D}^{*\mu\dagger}_a\bar{ D}^{*\nu}_b)(\partial_{\mu}V_{\nu}-\partial_{\nu}V_{\mu})_{ab},\nonumber \\  \\
\nonumber{\cal L}_{D^*DV} &=& \\
&&\!\!\!\!\!\!\!\!\!\!\!\!\!\!\!\!\!\!\!\!\!\!\!\!\!\!\!\!\!\!\!\!\!\!ig_{ D^{*}DV}\varepsilon_{\lambda\alpha\beta\mu}( D_b {\stackrel{\leftrightarrow \lambda}{\partial}} D^{*\mu\dag}_a
  +D_b^{*\mu}{\stackrel{\leftrightarrow \lambda}{\partial}} D^\dag_a )
  (\partial^\alpha{}V^\beta)_{ba} \nonumber \\
 &&\!\!\!\!\!\!\!\!\!\!\!\!\!\!\!\!\!\!\!\!\!\!\!\!\!\!\!\!\!\!\!\!\!+ig_{ {\bar D}^{*}\bar{D}V}\varepsilon_{\lambda\alpha\beta\mu}( \bar{D}_b {\stackrel{\leftrightarrow \lambda}{\partial}} \bar{D}^{*\mu\dag}_a
  +\bar{D}_b^{*\mu}{\stackrel{\leftrightarrow \lambda}{\partial}} \bar{D}^\dag_a )
  (\partial^\alpha{}V^\beta)_{ba}. \quad \quad
\end{eqnarray}
The matrix ${\cal M}$ contains
$\pi$, $K$, $\eta$ fields, which is  a $3\times 3$ hermitian and
traceless matrix. $V_{\mu}$ is analogous to ${\cal M}$
and contains $\rho$, $K^{*}$, $\omega$ and $\phi$. The matrices ${\cal M}$ and $V_{\mu}$ are expressed as
\begin{equation}\nonumber 
\label{5}{\cal M}=\left(
\begin{array}{ccc}
\frac{\pi^0}{\sqrt{2}}+\frac{\eta}{\sqrt{6}}&\pi^{+}&K^{+}\\
\pi^{-}&-\frac{\pi^{0}}{\sqrt{2}}+\frac{\eta}{\sqrt{6}}&K^{0}\\
K^{-}&\overline{K}^{0}&-\sqrt{\frac{2}{3}}\,\eta
\end{array}
\right), 
\end{equation}
\begin{equation}
V=\left(\begin{array}{ccc}
\frac{\rho^0}{\sqrt{2}}+\frac{\omega}{\sqrt{2}}&\rho^{+}&K^{*+}\\
\rho^{-}&-\frac{\rho^0}{\sqrt{2}}+\frac{\omega}{\sqrt{2}}&K^{*0}\\
K^{*-}&\overline{K}^{*0}&\phi
\end{array}
\right).
\end{equation}
The coupling constants are as follows,
\begin{eqnarray}
\nonumber g_{DD^{*}P}&=&-g_{\bar{D}\,\bar{D}^{*}P}=-\frac{2g}{f_{\pi}}\sqrt{m_{ D}m_{ D^{*}}}, \\
\nonumber g_{D^*D^{*}P}&=&g_{\bar{D}^*\,\bar{D}^{*}P}=-\frac{g}{f_{\pi}}, \\
\nonumber g_{{ DD}V}&=&-g_{\bar{ D}\,\bar{
D}V}=\frac{1}{\sqrt{2}}\beta g_{V}, \\
\nonumber g_{{ D^{*}D^{*}}V}&=&-g_{{\bar{
D}^{*}\bar{ D}^{*}}V}=-\frac{1}{\sqrt{2}}\,\beta g_{V}, \\
\nonumber g'_{{ D^{*}D^{*}}V}&=&-g'_{{\bar{
D}^{*}\bar{ D}^{*}}V}=-\sqrt{2}\,\lambda g_{V}{
m_{D^{*}}}, \\
g_{{ D D^{*}}V}&=&g_{{ \bar{D} \bar{D}^{*}}V}=\sqrt{2}\,\lambda g_{V}.
\end{eqnarray}
The effective Lagrangians between
$\sigma$ and heavy mesons are 
\begin{eqnarray}
{\cal L}_{DD\sigma}&=&g_{ DD\sigma}\,{ D}_a{
D}^{\dagger}_a\sigma+g_{\bar{D}\,\bar{D}\sigma}\,\bar{
D}_a\bar{ D}^{\dagger}_a\sigma, \\
\nonumber {\cal L}_{ {D}^{*}{D}^{*}{\sigma}}&=&g_{
D^{*}D^{*}\sigma}\, { D}^{*\mu}_a{ D}^{*\dagger}_{a\mu}\sigma+
g_{\bar{D}^{\,*}\bar{D}^{\,*}\sigma}\,
\bar{ D}^{*\mu}_a\bar{D}^{*\dagger}_{a\mu}\sigma,\\
\end{eqnarray}
and the relevant coupling constants are
\begin{eqnarray}
\nonumber g_{DD\sigma}&=&g_{
\bar{D}\,\bar{D}\,\sigma}=-2g_{S}{m_D},\\ g_{D^{*}D^{*}\sigma}&=&g_{\bar{D}^{*}\bar{D}^{*}\sigma}=2g_{S}{m_{D^{*}}}.
\end{eqnarray}

The effective Lagrangians depicting the charmed baryons with the light mesons read \cite{Yan:1992gz,Liu:2011xc},
\begin{eqnarray}
{\cal L}_{{B}_6{B}_6{P}}
&=&-\frac{g_1}{4m_{B_6}f_\pi }~\epsilon^{\alpha\beta\lambda\kappa}
    \langle\bar{{B}}_6{\overleftrightarrow{\partial}}_\kappa\gamma_\alpha \gamma_\lambda
    \partial_\beta\mathcal{M} ~{B}_6\rangle, \nonumber \\ \\
\nonumber {\cal L}_{{B}_6{B}_6{V}} &=& \frac{-i\beta_S
g_{V}}{2\sqrt{2}m_{B_6}}~\langle\bar{{B}}_6~ \overleftrightarrow{\partial}\cdot
{V}~ {B}_6\rangle \\
&&-\frac{i\lambda_S g_{V}}{3\sqrt{2}}~\langle\bar{{B}}_6\gamma_\mu
\gamma_\nu
(\partial^\mu {V}^\nu-\partial^\nu{V}^\mu){B}_6\rangle,\\
 {\cal L}_{{B}_6{B}_6\sigma} &=&
-\ell_S\langle\bar{{B}}_6~\sigma~{B}_6\rangle,\\ 
{\cal L}_{{B}_{\bar{3}}{B}_{\bar{3}}{V}} &=& \frac{i\beta_B
g_{V}}{2\sqrt{2}m_{{B}_{\bar{3}}}}~\langle\bar{{B}}_{\bar{3}}~ \overleftrightarrow{\partial}\cdot
{V}~ {B}_{\bar{3}}\rangle, \\
{\cal L}_{{B}_{\bar{3}}{B}_{\bar{3}}\sigma} &=&
\ell_B \langle\bar{{B}}_{\bar{3}}~\sigma~{B}_{\bar{3}}\rangle,\\
\nonumber \mathcal{L}_{{B}_{\bar{3}}{B}_6^{}{V}} &=&
       \frac{-i\lambda_Ig_V}{2\sqrt{6}\sqrt{m_{B_6}m_{{B}_{\bar{3}}}}}\varepsilon^{\mu\nu\lambda\kappa}
       \\ 
    &&\langle \bar{{B}}_6 \overleftrightarrow{\partial}_\mu \gamma^5\gamma_{\nu}\left(\partial_{\lambda} {V}_{\kappa}-\partial_{\kappa} {V}_{\lambda}\right){B}_{\bar{3}}\rangle +h.c.,
\end{eqnarray}
\begin{eqnarray}
\mathcal{L}_{{B}_{\bar{3}}{B}_6^{} {P}} &=& -\sqrt{\frac{1}{3}}\frac{g_4}{f_{\pi}}\langle\bar{{B}}_6\gamma^5\left(\gamma^{\mu}+v^{\mu}\right)\partial_{\mu}\mathcal{M}{B}_{\bar{3}}\rangle
    +h.c., \nonumber \\
\end{eqnarray}
where the partial $\overleftrightarrow{\partial}$ operates on the initial and final baryons. The heavy baryons matrices are defined as
\begin{equation}
{B}_{\bar{3}} = \left(\begin{array}{ccc}
        0\,    &\Lambda_c^+ \,     &\Xi_c^+ \,\\
        -\Lambda_c^+\,       &0\,      &\Xi_c^0\,\\
        -\Xi_c^+\,      &-\Xi_c^0\,     &0\,
\end{array}\right),
\end{equation} 
\begin{equation}
{B}_6^{} = \left(\begin{array}{ccc}
         \Sigma_c^{{}++} \,                 &\frac{1}{\sqrt{2}}  {\Sigma_c^{{}+}}\,   &\frac{1}{\sqrt{2}}{\Xi_c'^{+}}\,\\ 
         \frac{1}{\sqrt{2}}{\Sigma_c^{{}+}}\,      &\Sigma_c^{{}0}\,    &\frac{1}{\sqrt{2}}{\Xi_c'^{0}}\,\\
         \frac{1}{\sqrt{2}}{\Xi_c'^{+}}\,    &\frac{1}{\sqrt{2}}{\Xi_c'^{0}}\,      &\Omega_c^{0}\,
\end{array}\right).
\end{equation}
We list the model parameters used in this 
work in Table~\ref{Tab:InputCoup}, which are given in Refs.~\cite{Falk:1992cx,Isola:2003fh,Liu:2011xc}.

\begin{table*}
\caption{The values of the coupling constants adopted in our
calculation are taken from Refs.~\cite{Falk:1992cx,Isola:2003fh,Liu:2011xc}. The pion decay constant is $f_{\pi}=0.132$~GeV. The $\lambda$, $\lambda_S$, $\lambda_I$ are in the unit of GeV$^{-1}$. Others are in the unit of $1$.
}
\label{Tab:InputCoup}
\setlength{\tabcolsep}{3mm}{
\begin{tabular}{cccccccccccccccccccccc}\hline
$\beta$&$g$&$g_V$&$\lambda$ &$g_{_S}$&$\beta_S$&$\beta_B$&$\ell_S$& $\ell_B$&$g_1$&$g_4$&$\lambda_S$&$\lambda_I$\\
\hline
0.9\,&0.57\,&5.8\,&0.56\, &0.76\,&-1.74\,&$-\beta_S/2$\,&6.2\,&$-\ell_S/2$\,&-0.94\,&$3g_1/2\sqrt{2}$\,&-3.31\,&$-\lambda_S/2\sqrt{2}$\,\\
\hline
\end{tabular}}
\end{table*}
\begin{table*}
\caption{The $DD\psi$ and $D^*D^*\psi$ coupling constants were derived using the VMD model in Refs.~\cite{Lin:1999ad,Sibirtsev:2000aw,Liu:2001ce,Oh:2007ej}. To get the $DD^*\psi$ coupling, we have used the relation in the heavy quark mass limit, the $\bar{M}_{D}$ is the average mass scale of the $D$ and $D^{*}$ mesons~\cite{Oh:2007ej,Deandrea:2003pv}. Following the $SU(4)$ relations from Refs.~\cite{Liu:2001ce,Dong:2009tg},
we can get $g_{D^{(*)}N\Lambda_c}=g_{K^{(*)}N\Lambda}$, 
$g_{D^{(*)}N\Sigma_c}=g_{K^{(*)}N\Sigma}$, $\kappa_{D^*N\Sigma_c}=\kappa_{K^*N\Sigma }$, $\kappa_{D^*N\Lambda_c}=\kappa_{K^*N\Lambda}$. The values of the coupling constants used in this work are taken from Refs.~\cite{Rijken:1998yy,Kim:2011rm,Oh:2006in}.
}
\label{Tab:InputCoup2}
\setlength{\tabcolsep}{3mm}{
\begin{tabular}{ccccccccc}\hline
$g_{DD\psi}$&$g_{D^*D^*\psi}$&$g_{DD^*\psi}$&$g_{DN\Lambda_c}$ &$g_{D^*N\Lambda_c}$&$g_{DN\Sigma_c}$&$g_{D^*N\Sigma_c}$ &$\kappa_{D^*N\Sigma_c}$ & $\kappa_{D^*N\Lambda_c }$ \\
\hline
7.64\,&7.64\,&$g_{DD\psi}/\bar{M}_D$\,&$-13.24$\, &$-4.26$\,&3.58\,&-2.46 & -0.47 & 2.66 \,\\
\hline
\end{tabular}}
\end{table*}
For the transitions of the $\Lambda_c \bar{D}^{(*)}$ and $\Sigma_c \bar{D}^{(*)}$ channels to the lowest channel $J/\psi p$, we 
use effective Lagrangians~\cite{Lin:1999ad,Sibirtsev:2000aw,Liu:2001ce,Oh:2007ej}, 
\begin{eqnarray}
\mathcal{L}_{DD\psi}&=&-ig_{DD\psi} \psi^{\mu} (\partial_{\mu}D\bar{D}-D\partial_{\mu}\bar{D}), \\
\mathcal{L}_{DD^*\psi}&=&-g_{DD^*\psi} \epsilon^{\mu\nu\alpha\beta}\partial_{\mu} \psi_{\nu} (\partial_{\alpha}D^*_{\beta}\bar{D}+D\partial_{\alpha}\bar{D}^*_{\beta}),\,\,\,\,\,\,\,\, \\
\nonumber \mathcal{L}_{D^*D^*\psi}&=&-ig_{D^*D^*\psi}\Big( \psi^{\mu} (\partial_{\mu}D^{*\nu}\bar{D}^*_{\nu}-D^{*\nu}\partial_{\mu}\bar{D}_{\nu}^*)  \\
&&+\psi^{\nu} (D^{*\mu}\partial_{\mu}\bar{D}_{\nu}^*-\partial_{\mu}D^*_{\nu}\bar{D}^{*\mu}) \Big), \\
\mathcal{L}_{DN\Sigma_c}&=&-\frac{g_{DN\Sigma_c}}{m_N+m_{\Sigma_c}}\bar{N} \gamma_5 \gamma_{\mu} \partial^{\mu}\bar{D} \Sigma_c +h.c., \\
\mathcal{L}_{D^*N\Sigma_c}&=&-
\nonumber g_{D^*N\Sigma_c}\Big( \bar{N} \gamma_{\mu} \bar{D}^{*\mu} -  \frac{\kappa_{D^*N\Sigma_c}}{2m_N} \sigma_{\mu\nu}\partial^{\nu}\bar{D}^{*\mu}  \Big) \\
&&+h.c..
\end{eqnarray}
$\mathcal{L}_{D^{(*)}N\Lambda_c}$ has a same structure as $\mathcal{L}_{D^{(*)}N\Sigma_c}$.
The involved coupling constants used in this work are given in 
Table~\ref{Tab:InputCoup2}.

The effective Lagrangians for the electromagnetic interactions for the mesons and baryons are given as~\cite{Kim:2011rm,Oh:2006in}
\begin{eqnarray}
\mathcal{L}_{\gamma D D^*}&=&g_{\gamma DD^{*}} \epsilon^{\mu\nu\alpha\beta} (\partial_{\mu}A_{\nu})(\partial_{\alpha}D_{\beta}^{*})D + h.c., \\
\mathcal{L}_{\gamma D D}&=& ie_{\bar{D}} A_{\mu}\Big( \partial^{\mu}\bar{D} D -\partial^{\mu}{D}\bar{D}     \Big), \\
\nonumber \mathcal{L}_{\gamma D^*D^*}&=&-ie_{\bar{D}^*}\Big( A^{\mu} (\partial_{\mu}D^{*\nu}\bar{D}^*_{\nu}-D^{*\nu}\partial_{\mu}\bar{D}_{\nu}^*)  \\ 
&&+A^{\nu} (D^{*\mu}\partial_{\mu}\bar{D}_{\nu}^*-\partial_{\mu}D^*_{\nu}\bar{D}^{*\mu}) \Big), \\
\mathcal{L}_{\gamma NN}&=&-\bar{N}\Big(e_N \gamma_{\mu}-\frac{e\kappa_N}{2m_N} \sigma_{\mu\nu}\partial^{\nu}   \Big) A^{\mu} N,\\ 
\mathcal{L}_{\gamma \Sigma_c\Sigma_c}&=&-\bar{\Sigma}_c\Big(e_{\Sigma_c} \gamma_{\mu}-\frac{e\kappa_{\Sigma_c}}{2m_{\Sigma_c}} \sigma_{\mu\nu}\partial^{\nu}   \Big) A^{\mu} \Sigma_c.
\end{eqnarray}
$\mathcal{L}_{\gamma \Lambda_c\Lambda_c}$ has a same structure as $\mathcal{L}_{\gamma \Sigma_c\Sigma_c}$.
The electriccharge of the $\bar{D}^{(*)}$ meson is denoted as $e_{\bar{D}^{(*)}}$. The coupling constants $g_{\gamma DD^*}$ can be calculated from the experimental data for $\Gamma_{D^{*\pm}\to D^{\pm} \gamma}=1.33$ keV~\cite{ParticleDataGroup:2024cfk} and $\Gamma_{D^{*0}\to D^{0} \gamma}=21.24$ keV~\cite{Rosner:2013sha}, which gives $g_{D^{*\pm}D^{\pm}\gamma}=0.141~{\rm GeV^{-1}}$ and $g_{D^{*0}D^{0}\gamma}=-0.558~{\rm GeV^{-1}}$.   

The $\kappa_{N}$, $\kappa_{\Lambda_c}$ and $\kappa_{\Sigma_c}$ are the anomalous magnetic moments of the $N$, $\Lambda_c$ and $\Sigma_c$ baryons, respectively. For the anomalous magnetic moment of the proton, we taken
$\kappa_{p}=1.79$~\cite{ParticleDataGroup:2024cfk}.
The $\kappa_{\Lambda_c}$ can be related to the magnetic moment $\mu_{\Lambda_c}$ by
\begin{align}
\mu_{\Lambda_c}=\frac{(e_{\Lambda_c}+e\kappa_{\Lambda_c})}{2m_{\Lambda_c}}.
\end{align}
This is analogous for the $\Sigma_c$. Using the relations given in Ref.~\cite{Fomin:2019wuw}, 
\begin{align}
\mu_{\Lambda_c}=0.4\mu_N, \quad \mu_{\Sigma_c^{++}}=2.54\mu_N, \nonumber \\ \mu_{\Sigma_c^{+}}=0.54\mu_N  \quad  \mu_{\Sigma_c^{0}}=-1.46\mu_N, 
\end{align}
with $\mu_N=\frac{e}{2m_p}$, then one get
\begin{align}
\kappa_{\Lambda_c}=-0.025, \quad \kappa_{\Sigma_c^{++}}=4.64, \nonumber \\ \kappa_{\Sigma_c^{+}}=0.41, \quad  \kappa_{\Sigma_c^{0}}=-3.81.
\end{align}
In this work, we use the isospin phase conventions,
\begin{align}
 &|D^+\rangle =|\frac{1}{2},\frac{1}{2}\rangle,\quad |D^0\rangle =-|\frac{1}{2},-\frac{1}{2}\rangle, \\
 &|\bar{D}^0\rangle =|\frac{1}{2},\frac{1}{2}\rangle,\quad |D^-\rangle =|\frac{1}{2},-\frac{1}{2}\rangle,\\
 &|p\rangle =|\frac{1}{2},\frac{1}{2}\rangle,\quad |n\rangle =|\frac{1}{2},-\frac{1}{2}\rangle, \\
 &|\Lambda_c\rangle =|0,0\rangle,\\
 &|\Sigma_c^{++}\rangle= |1,+1\rangle, \quad  |\Sigma_c^{+}\rangle= |1,0\rangle,  \quad  |\Sigma_c^{0}\rangle= |1,-1\rangle.
\end{align}
The isospin states $D^*$ are same as those of $D$.
In this work, only particle pairs with isopsin $I=\frac{1}{2}$, which relates to $J/\psi p$ final state, will be considered. The wave function with isopsin $I=\frac{1}{2}$
can be written as
\begin{align}
&|\frac{1}{2},+\frac{1}{2}\rangle = \sqrt{\frac{2}{3}}|\Sigma_c^{++}D^{(*)-}-\sqrt{\frac{1}{3}}|\Sigma_c^{+}\bar{D}^{(*)0}, \\
&|\frac{1}{2},+\frac{1}{2}\rangle = |J/\psi p\rangle, \\
&|\frac{1}{2},+\frac{1}{2}\rangle = |\Lambda_c^{+}\bar{D}^{(*)0}\rangle.
\end{align}
Then we can get the isospin factors $IF$ for every exchange process corresponding to $I=\frac{1}{2}$ as given in 
Table~\ref{Tab:IsospinNum}.
  
\begin{table}
\caption{The isospin factors $IF$ for every exchange process corresponding to $I=\frac{1}{2}$.
}
\label{Tab:IsospinNum}.
\setlength{\tabcolsep}{3mm}{
\begin{tabular}{cccccc}\hline
   & $\pi$ &$\eta$&$\rho$ &$\omega$ &$\sigma$\\
\hline
$\Sigma_c \bar{D}^{(*)}\to \Sigma_c \bar{D}^{(*)}$ & -1 &$\frac{1}{6}$\,&-1&$\frac{1}{2}$&1\\
$\Lambda_c\bar{D}^{(*)}\to\Lambda_c \bar{D}^{(*)}$ & 0 &              0 &0 & 1            &2\\
$\Lambda_c\bar{D}^{(*)}\to\Sigma_c \bar{D}^{(*)}$ & $-\frac{\sqrt{6}}{2}$ &              0 &$-\frac{\sqrt{6}}{2}$ & 0            &0\\  \\
  & $D$ &$D^*$& & &\\
$\Sigma_c \bar{D}^{(*)}\to J/\psi p $ &$\sqrt{3}$ & $\sqrt{3}$  &   &   &\\
$\Lambda_c\bar{D}^{(*)}\to  J/\psi p$ & 1          & 1          &   &   &\\
\hline
\end{tabular}}
\end{table}

\begin{table*}
\caption{Pole positions $\sqrt{s_p}$ (GeV) and their couplings to different channels $g_i$ (GeV). The parameters $\alpha$ listed in the table correspond to light meson $\pi$, $\eta$, $\sigma$, $\rho$ and $\omega$ exchanges. For heavy meson $D$ and $D^{*}$ exchanges, the values of $\alpha$ are fixed to $0.7$ throughout this work.
}
\label{Tab:PopPole}.
\setlength{\tabcolsep}{3mm}{
\begin{tabular}{cccccccc}\hline
 $\alpha$  & $J^{P}$ &$\sqrt{s_p}$&  $g_{\Lambda_c\bar{D}}$ & $g_{\Lambda_c\bar{D}^*}$&  $g_{\Sigma_c\bar{D}}$  &  $g_{\Sigma_c\bar{D}^*}$ & $g_{J/\psi p}$ \\
\multirow{2}*{$1.6$}
  & $\frac{1}{2}^-$ &$4.3175-i0.0032$& $1.58-i0.26$  & $-1.06-i0.05$ & $11.76+i1.26$ & $-1.53-i7.89$  & $-0.12+i0.11$ \\
  & $\frac{1}{2}^-$ &$4.4315-i0.0140$& $3.26-i0.43$ & $-0.39+i0.11$& $-0.94+0.90i$  & $24.56+i1.24$   & $-0.56+i0$\\
$2.1$  & $\frac{3}{2}^-$ &$4.4562-i0.0025$& $0+i0$  & $-1.16+i0.27$ & $0+i0$ & $12.57+i0.36$  & $-0.15+i0$\\
\hline
\end{tabular}}
\end{table*}

With the help of the effective Lagrangians, the constructed $t$-channel $\pi$, $\eta$, $\sigma$, $\rho$ and $\omega$ exchange potentials of the $\Lambda_c \bar{D}^{(*)}$, $\Sigma_c \bar{D}^{(*)}$ and $J/\psi$ coupled-channel interactions are given in Appendix~\ref{sec:exppotenA}. 
For the $\gamma p \to \Sigma_c \bar{D}^{(*)}$ and $\Lambda_c \bar{D}^{(*)}$ transition potentials, the $t$-channel $\bar{D}$ and $\bar{D}^*$ exchanges, $s$-channel nucleon exchange, $u$-channel $\Sigma_c$ exchange and a contact term are considered. The explicit formulas of the potentials are given
in Appendix~\ref{sec:exppotenB} and~\ref{sec:exppotenC}.

\subsection{Form factors and gauge invariance}
To parametrize short-distance dynamics,
the following form factors are introduced for $t$, $s$ and $u$-channel exchange processes,
\begin{align}
\label{eq:form}
f_{t}=\Big(\frac{\Lambda^2-m^2}{\Lambda^2-q_t^2}\Big), \quad f_{u/s}=\frac{\Lambda^4}{\Lambda^4+(q_{u/s}^2-m^2)^2}.
\end{align}
Here, $m$ denote the masses of the exchanged particles, $\Lambda$ are the cutoff parameters. 

The four-momenta of the exchanged particles for $t$, $u$ and $s$-channel are defined, respectively, as 
\begin{align}
\label{eq:Tranmome}
q_t=k_3-k_1, \quad q_u=k_4-k_1,\quad q_s=k_1+k_2.
\end{align}
  
The gauge invariance or electromagnetic current conservation is a fundamental requirement which any physical theory should fulfill.
In present work, the contact terms are introduced by the minimal gauge substitution $\partial_{\mu} \to \partial_{\mu} + ie_{D^{(*)}}A_{\mu}$ to the $D^{(*)}N\Lambda_c$ and $D^{(*)}N\Sigma_c$ interactions~\cite{Schwartz:2014sze,Mosel1999}.
For the vertices derived through minimal substitution, the invariance of the theory under local gauge transformations can be tested by the replacement $\epsilon^{\mu}(k_1,\lambda_1) \to k_1^{\mu}$ in the amplitudes given in Appendix~\ref{sec:exppotenB} and~\ref{sec:exppotenC}.
The sum of the $s$,
$t$ and $u$-channel potentials and contact term fulfills the constraint
\begin{eqnarray}
\label{eq:gaugeinv}
\sum_{i=s,u,t,C}{\Gamma}_{\mu}k_1^{\mu}=0,
\end{eqnarray}
which is shown in Appendix~\ref{sec:exppotenB} and~\ref{sec:exppotenC}.

However, the inclusion of the form factors will violate the gauge invariance.
The form factors have different values for each term, all terms in Eq.~\eqref{eq:gaugeinv} are weighted differently once form factors are included. 
In order to restore gauge invariance in this situation, many schemes were
proposed as discussed in Refs.~\cite{Gross:1987bu,Ohta:1989ji,vanAntwerpen:1994vh,Haberzettl:1997jg,Haberzettl:1998aqi,Kvinikhidze:1998xn,Davidson:2001rk,Haberzettl:2006bn,Huang:2011as}. 
There is no unique way to maintain gauge invariance once form factors have been introduced.
In present work, we use the scheme proposed in Ref.~\cite{Davidson:2001rk}, 
which repects the crossing symmetry and ensures that the additional contributions are free of poles. In this scheme, the hadronic form factors in Eq.~\eqref{eq:form} multiplying the charge contributions of the $\gamma p \to \Sigma_c^{++} D^{(*)-}$ transitions given in Appendix~\ref{sec:exppotenB} and~\ref{sec:exppotenC} are replaced by a common form factor
\begin{align}
\label{eq:comfac}
\nonumber \hat{f}_{t,s,u}=&f_{t}+f_{s}+f_{u}-f_{t}f_{s}-f_{t}f_{u} \\
&-f_{s}f_{u}+f_{t}f_{s}f_{u}.
\end{align}
Here, for the $\gamma p \to \Sigma_c^{++} D^-$ and $\Sigma_c^{++} D^{*-}$ potentials, the $t$-channel exchanged mesons are $D^-$ and $D^{*-}$, respectively.

For the $\gamma p \to \Sigma_c^{+} \bar{D}^{(*)0}$ and $\Lambda_c^{+} \bar{D}^{(*)0}$ transition, the form factor in Eq.~\eqref{eq:comfac} is 
applied by setting $f_t=0$, since the corresponding potentials are absent.

\begin{figure}[tbhp]
\begin{center}
\includegraphics [scale=0.68] {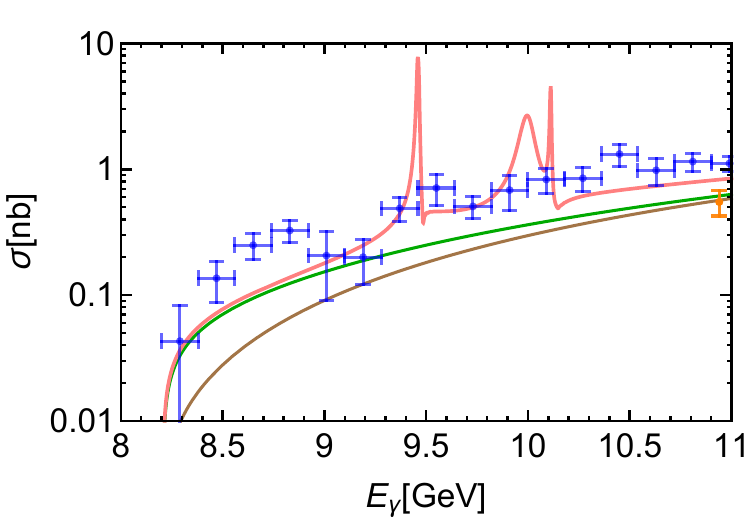}
\caption{The total cross section for $\gamma p \to J/\psi p$ reaction as a function of the
photon laboratory energy $E_{\gamma}$. The brown line stands for the Pomeron exchange. The green line stands for the contribution with both Pomeron and $\sigma$ exchanges. The red line represents the full result including Pomeron exchange, $\sigma$ exchange and open-charm state rescattering. The blue circles indicate the GlueX
data~\cite{GlueX:2023pev}. The orange
circle indicates the Cornell
data~\cite{Gittelman:1975ix}.
}
\label{fig:GluexData}
\end{center}
\end{figure}

\section{Numerical results and discussions}
\label{sec:Results}
\subsection{Poles from the coupled-channel interactions}

The $P_c$ signals manifest as pole singularities in the 
$\Lambda_c \bar{D}^{(*)}$, $\Sigma_c \bar{D}^{(*)}$
and $J/\psi p$ scattering amplitudes. The effective couplings $g_i$ of these poles to channel $i$ can be obtained from the
calculated residues of the amplitude ${T}(s,p',p)$ given in Eq.~\eqref{eq:bsess} at the pole positions,
\begin{eqnarray}
\label{eq:EfeC}
g_{i'}g_{i}= \mathop{\rm lim}_{s\to s_{p}}\frac{1}{4\pi} (s-s_{p})T_{i'i}\,,
\end{eqnarray}
where $\sqrt{s_p}$ are the pole positions.

\begin{figure*}[tbhp]
\begin{center}
\includegraphics [scale=0.45] {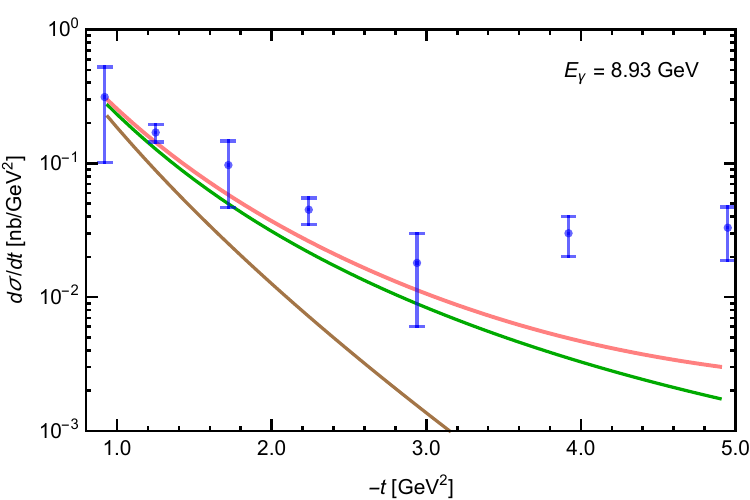}
\includegraphics [scale=0.45] {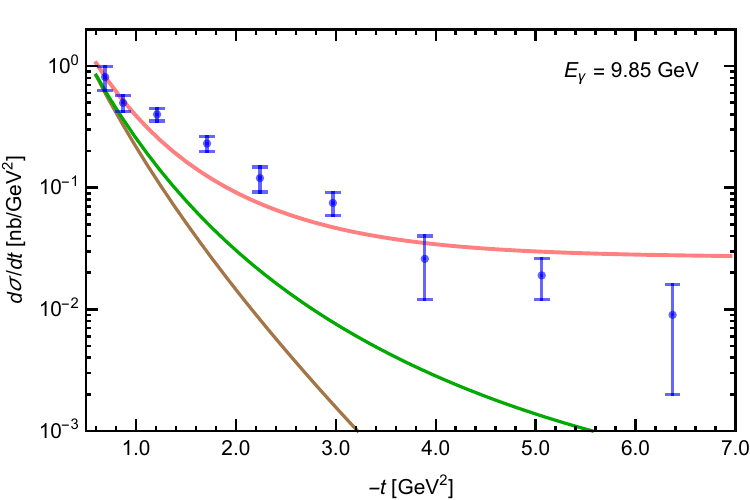}
\includegraphics [scale=0.45] {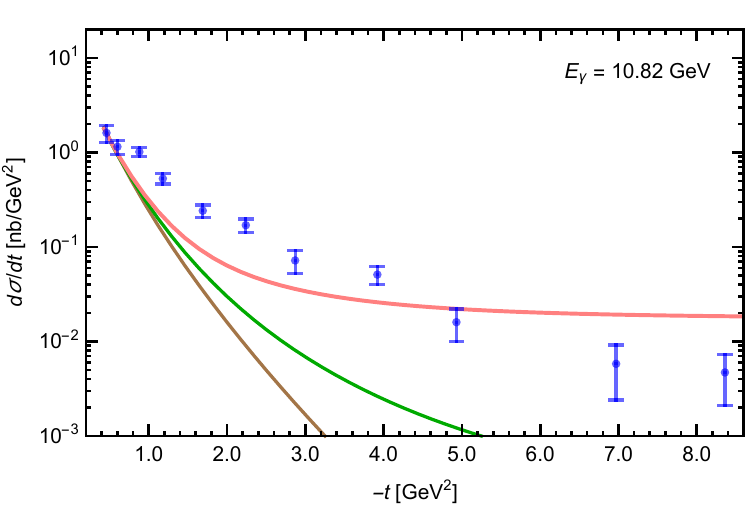}
\caption{The differential cross sections for $\gamma p \to J/\psi p$ reaction. The line notations are same as those in Fig.~\ref{fig:GluexData}. The experimental data are taken from GlueX Collaboration~\cite{GlueX:2023pev}.}
\label{fig:GluexDatadiff}
\end{center}
\end{figure*}
\begin{figure*}[tbhp]
\begin{center}
\includegraphics [scale=0.45] {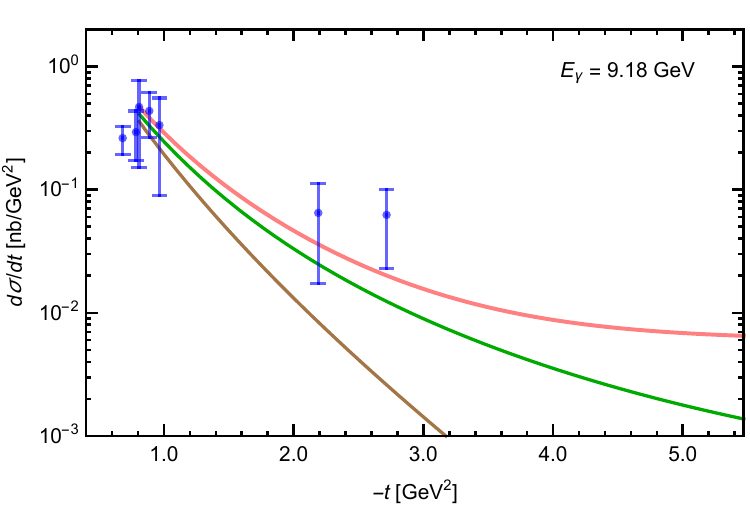}
\includegraphics [scale=0.45] {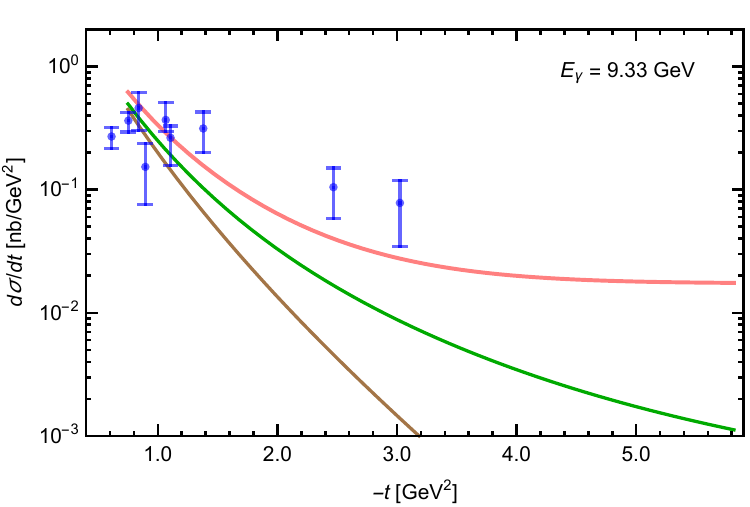}
\includegraphics [scale=0.45] {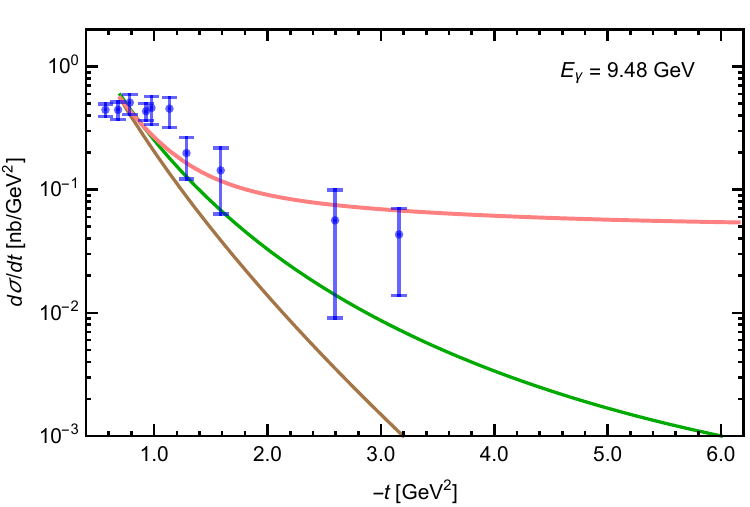}
\includegraphics [scale=0.45] {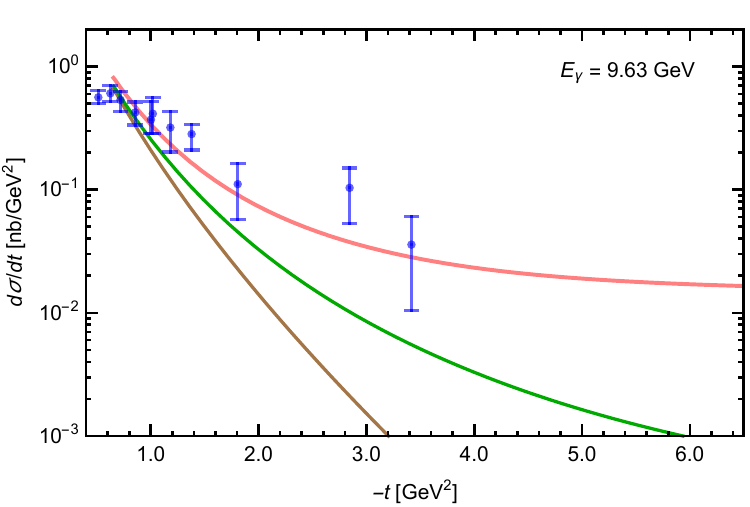}
\includegraphics [scale=0.45] {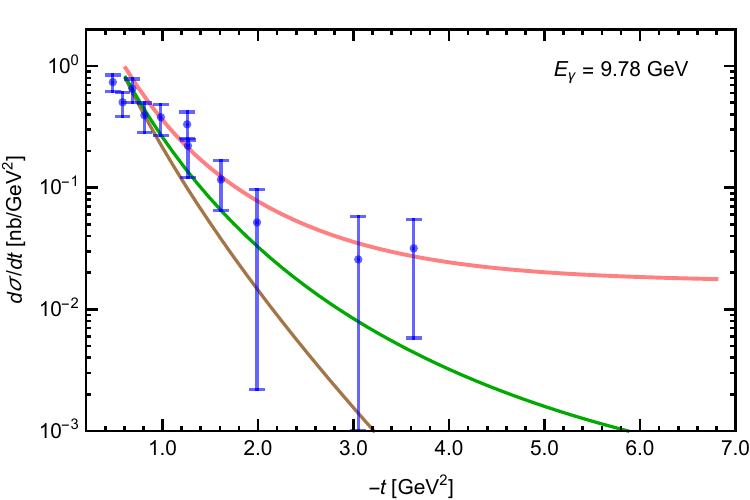}
\includegraphics [scale=0.45] {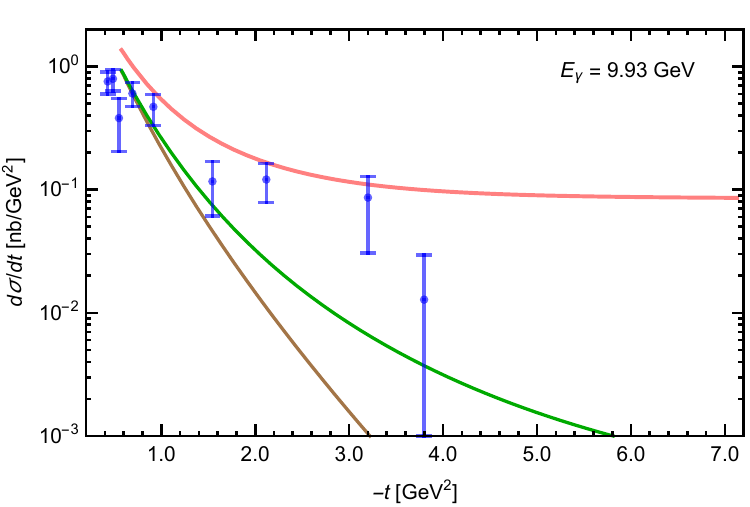}
\includegraphics [scale=0.45] {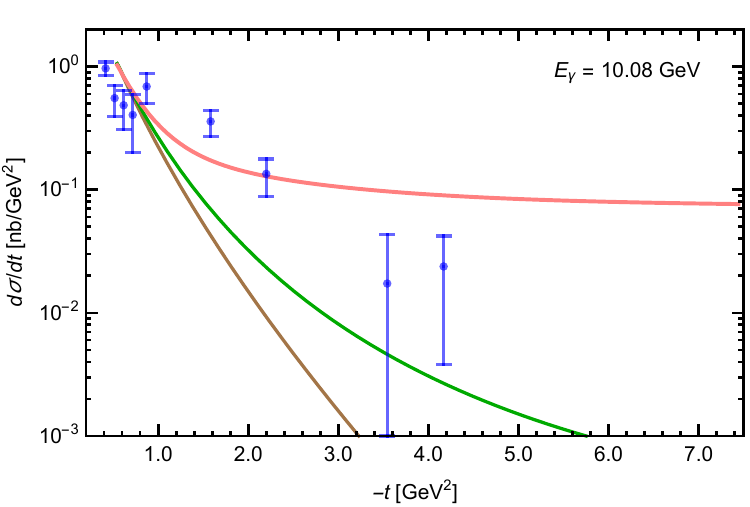}
\includegraphics [scale=0.45] {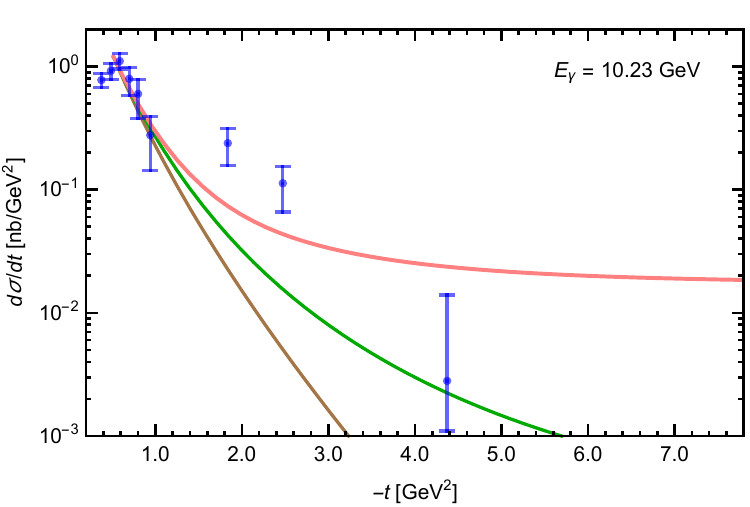}
\includegraphics [scale=0.45] {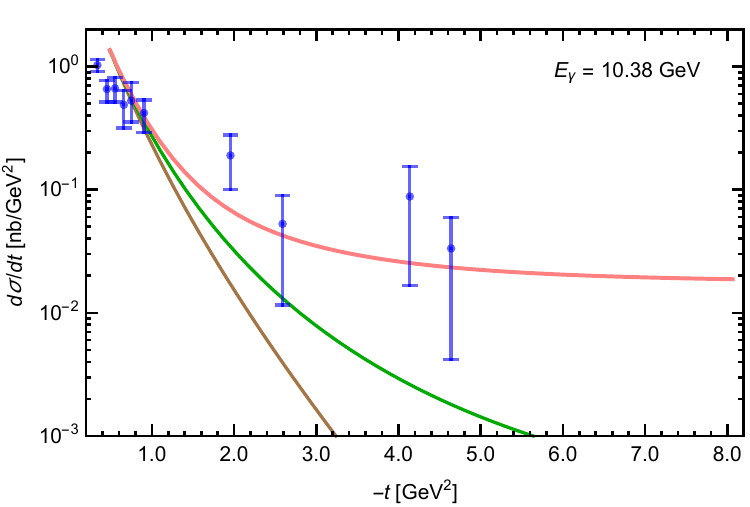}
\includegraphics [scale=0.45] {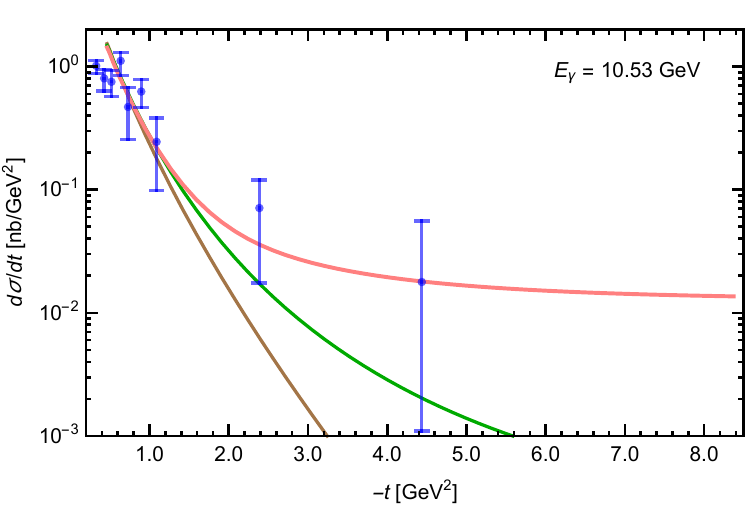}
\caption{The differential cross sections for $\gamma p \to J/\psi p$ reaction. The line notations are same as those in Fig.~\ref{fig:GluexData}. The experimental data are taken from $J/\psi$-$007$ experiment~\cite{Duran:2022xag}.}
\label{fig:psi007diff}
\end{center}
\end{figure*}

For $t$-channel exchange, the parameter in Eq.~\eqref{eq:form}
is taken to be
$\Lambda=m+\alpha\Lambda_{\rm QCD}$ with $\Lambda_{\rm QCD}=0.22$~GeV. 
For light meson $\pi$, $\eta$, $\sigma$, $\rho$ and $\omega$ exchanges, the values of $\alpha$ are adjusted to produce the poles with their positions approximately to the LHCb results~\cite{LHCb:2015yax,LHCb:2016ztz,LHCb:2019kea}. In present work, the values of $\alpha$ are chosen to be $1.6$ and $2.1$ for $J^{P}=\frac{1}{2}^-$ and $\frac{3}{2}^-$ channels, respectively. For heavy meson $D$ and $D^{*}$ exchanges, the value of $\alpha$ is fixed to $0.7$ throughout this work.
The scheme dynamically generates two states having a spin-parity of $\frac{1}{2}^-$ and one having $\frac{3}{2}^-$. The positions of
the poles in complex energy $\sqrt{s}$-plane and their couplings to different channels are listed in Table~\ref{Tab:PopPole}. 
The lowest pole has a spin-parity $J^P=\frac{1}{2}^-$ and is located on the physical Riemann sheets for the $\Sigma_c \bar{D}$ and $\Sigma_c \bar{D}^*$ channels
and the unphysical Riemann sheets for the $\Lambda_c \bar{D}$ and $\Lambda_c \bar{D}^{*}$ and $J/\psi p$ channels. 
The couplings of this pole to the $\Lambda_c \bar{D}$ and $\Lambda_c \bar{D}^{*}$ and $J/\psi p$ channels are weak, to $\Sigma_c \bar{D}$ and $\Sigma_c \bar{D}^*$ channels are stronger.
This pole has largest coupling to the $\Sigma_c \bar{D}$ channel and can be assigned to the $P_c(4312)$. 

The other two poles have spin-parity $J^P=\frac{1}{2}^-$ and $\frac{3}{2}^-$, respectively. They are located on the physical Riemann sheet for the $\Sigma_c \bar{D}^{*}$ channel and the unphysical Riemann sheets for the $\Lambda_c \bar{D}$, $\Lambda_c \bar{D}^{*}$ and $J/\psi p$ channels.
These two poles have dominant coupling to the $\Sigma_c \bar{D}^{*}$ channel and can be assigned to $P_c(4440)$ and $P_c(4457)$, respectively. The couplings of these two poles to $\Lambda_c \bar{D}$, $\Lambda_c \bar{D}^{*}$ and $J/\psi p$ channels are weak.

\subsection{$J/\psi$ photoproduction}

In this subsection, we present model results for the photon laboratory
energy $E_{\gamma}$ from $8.2$ GeV up to 11.0~GeV. In $\gamma p \to \Lambda_c \bar{D}^{(*)}$ and $\gamma p \to \Sigma_c \bar{D}^{(*)}$ transition potentials, the values of $\alpha$ corresponding to $t$-channel $D$ and $D^*$ exchanges 
are chosen same as those used before in calculating the coupled-channel interaction. For $\sigma$ exchange in $\gamma p \to J/\psi p$ reaction,
the value of $\alpha$ is chosen to be $2.1$, which is same as that used in calculating the coupled-channel interaction in $J^P=\frac{3}{2}^-$ channel.
For the $s$-channel nucleon and $u$-channel $\Sigma_c$ exchanges, we take the cutoff $\Lambda=0.5$~GeV, which is used to constrain $K^*\Lambda$ and $K^*\Sigma$ photoproductions~\cite{Kim:2011rm,Oh:2006in}. Fig.~\ref{fig:GluexData} shows the $E_{\gamma}$ dependence of the total cross sections for the $\gamma p \to J/\psi p$ reaction. Figs.~\ref{fig:GluexDatadiff} and~\ref{fig:psi007diff} show the differential cross sections $d\sigma/dt$ at different energies $E_{\gamma}$.
We find that with the parameters we used, 
the predicted total and differential cross sections are within the range of the available data. We will discuss about the total and differential cross sections in the following subsections.

\begin{figure*}[tbhp]
\begin{center}
\includegraphics [scale=0.6] {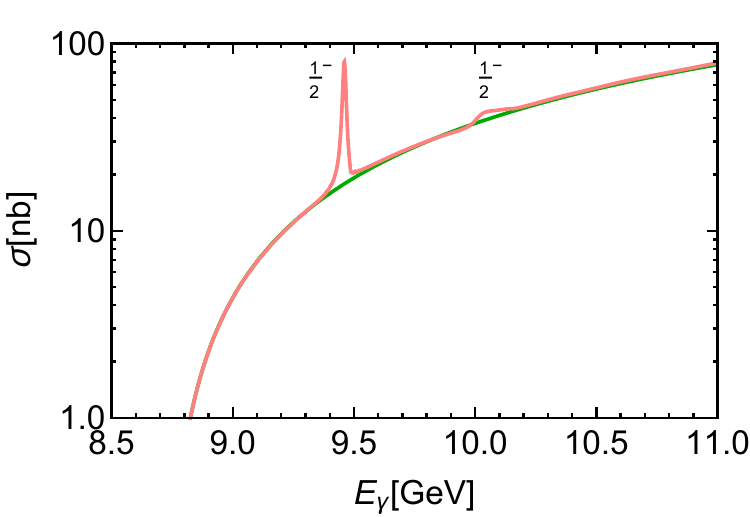}
\includegraphics [scale=0.6] {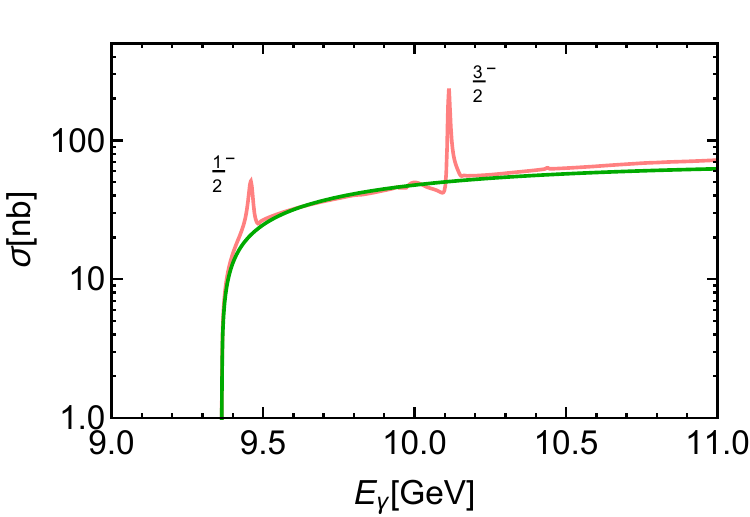}
\caption{The left and right panels show the total cross sections of $\gamma p \to \Lambda_c\bar{D}$ and $\gamma p \to \Lambda_c\bar{D}^*$ reactions, respectively. The green and red lines stand for the results without and with including open-charm state rescattering, respectively.}
\label{fig:OpenCharm}

\end{center}
\end{figure*}

\subsubsection{Total cross section}
In Fig.~\ref{fig:GluexData}, the brown line stands for the Pomeron exchange. The green line stands for the contribution with both Pomeron and $\sigma$ exchanges. The red line represents the full result including Pomeron exchange, $\sigma$ exchange and open-charm state rescattering.
One can see that in the very near threshold region, the Pomeron exchange contribution is weak,
and increases with increasing $E_{\gamma}$.
The $\sigma$ exchange is dominant in the 
very near threshold region, and decreases with increasing $E_{\gamma}$. It is observed that the total cross section rises sharply in the very near threshold region due to the $\sigma$ exchange. The open-charm state rescattering contribution gets enhanced in the regions of the $P_c(4312)$, $P_c(4440)$ and $P_c(4457)$ productions. The results indicate that the total cross section from pentaquark contributions is of order 1 nb. 
Outside the pentaquark production regions, the open-charm state rescattering contribution becomes small.

\subsubsection{Differential cross section}
The experimental data in Figs.~\ref{fig:GluexDatadiff} and~\ref{fig:psi007diff} are taken from the GlueX and $J/\psi$-$007$ experiments, respectively. We can see that at small momentum transfer $|t|$, the differential cross sections $d\sigma/dt$ are almost determined by the Pomeron exchange.
With the increase of the momentum transfer $|t|$, the Pomeron exchange drops exponentialy, the $\sigma$ exchange and the open-charm state rescattering
become dominant. Another feature that we can find is the enhancement in the predicted differential cross section at larger $|t|$
in the region of the $P_c(4312)$, $P_c(4440)$ and $P_c(4457)$ productions.

\subsection{Open-charm state production}
As it turns out in the present model, it is also possible to simultaneously predict the open-charm state photoproduction.
The total cross sections corresponding
to $\gamma p \to \Lambda_c\bar{D}$ and $\gamma p \to \Lambda_c\bar{D}^*$ reactions are shown in left and right panels of Fig.~\ref{fig:OpenCharm}, respectively. The green and red lines stand for the results without and with including open-charm state rescattering, respectively.
From the left panel of Fig.~\ref{fig:OpenCharm}, 
one can find that two resonance structures both with spin-parity $\frac{1}{2}^-$ show up in $\gamma p \to \Lambda_c\bar{D}$ reaction due to pentaquark contributions. Similar to the $J/\psi p$ production, the open-charm state rescattering contribution to the total cross sections is small except for the regions of pentaquark productions. There is no signal in this channel corresponding to pentaquark with spin-parity $\frac{3}{2}^-$, since we have only considered $S$-wave interactions.

From the right panel of Fig.~\ref{fig:OpenCharm}, 
one can find that the pentaquarks with spin-parity $\frac{1}{2}^-$ and $\frac{3}{2}^-$ produce two sharp peaks in 
$\gamma p \to \Lambda_c\bar{D}^*$ reaction at their pole positions. In this channel, only the lower one of the two pentaquarks, which have a spin-parity of $\frac{1}{2}^-$, shows up. And there is no significant resonance signal from the higher one due to its weak coupling to $\Lambda_c\bar{D}^*$ channel as shown in Table~\ref{Tab:PopPole}.

\section{Summary}
In this work, we have constructed a dynamical coupled-channel approach to photon-induced $J/\psi$ production to search for the trace of the pentaquarks $P_c$. The $J/\psi$ photoproduction mechanism is described by the Pomeron exchange, $\sigma$ exchange and open-charm state rescattering. The considered open-charm intermediate states are $\Lambda_c \bar{D}^{(*)}$ and $\Sigma_c \bar{D}^{(*)}$. Scattering amplitude involving $\Lambda_c \bar{D}^{(*)}$, $\Sigma_c \bar{D}^{(*)}$ and $J/\psi p$ interactions is calculated from solving a set of coupled-channel integral equations. The pentaquarks $P_c$ are dynamically generated from the coupled-channel interactions.

To reproduce the experimental data of the total cross section, the $\sigma$ exchange 
plays non-negligible contribution. One feature of the $\sigma$ exchange is that it dominants the total cross section in the very near threshold region, and its contribution decreases with increasing $E_{\gamma}$. The open-charm state rescattering contribution to the total cross sections is small except for the regions of the $P_c(4312)$, $P_c(4440)$ and $P_c(4457)$ productions. The results indicate that the $J/\psi$ photoproduction cross section from pentaquark contributions is of order 1 nb.

The predicted differential cross sections are compared to the experimental data from GlueX and $J/\psi$-$007$ experiments. Only the Pomeron exchange fails to reproduce the observed $t$ dependence of the differential cross section. The discrepancy can be removed by including the $\sigma$ exchange and open-charm state rescattering. At large momentum transfer $|t|$, the inclusion of the open-charm state rescattering will make the differential cross section enhanced to a certain extent in the regions of the $P_c(4312)$, $P_c(4440)$ and $P_c(4457)$ productions.

The emergence of the pentaquarks may break the QCD factorization in the pentaquark production regions. Outside these regions,
the open-charm state rescattering contribution is small.
Our results provide an important information to illustrate the near threshold $J/\psi$ photoproduction. 

\acknowledgments 

I wish to thank Feng-Kun Guo for careful reading of the manuscript and stimulating conversations. I also thank Jia-Jun Wu and Bing-Song Zou for helpful discussions. 
The results described in this paper are supported by HPC Cluster of ITP-CAS. This work is supported in part by the National Natural Science Foundation of China (NSFC) under Grants No. 12405106 and No. 12247139, by the Chinese Academy of Sciences under Grants No. XDB34030000 and No. YSBR-101; by the National Key R\&D Program of China under Grant No. 2023YFA1606703.

\appendix
\begin{widetext}
\section{The $\Lambda_c \bar{D}^{(*)}$, $\Sigma_c \bar{D}^{(*)}$ and $J/\psi p$ coupled-channel interaction potentials}
\label{sec:exppotenA}
The interaction potentials can be written as
\begin{align}
\langle \, \vec{p}\,'\lambda_3\lambda_4|V(s)| \vec{p}\,\lambda_1\lambda_2\rangle= IF \cdot f_{t}\cdot V,
\end{align}
where $IF$ and $f_{t}$ are the isospin factor and form factor for every exchange process. The isospin factor $IF$
for every exchange process is given in Table~\ref{Tab:IsospinNum}, the $t$-channel exchange form factor is given in Eq.~\eqref{eq:form}. The explicit form of the potential $V$ for every $t$-channel one-boson-exchange process is given as follows.

$\Sigma_c \bar{D} \to \Sigma_c \bar{D}$ interaction, 
\begin{align}
V_V=&-\frac{\beta g_V}{\sqrt{2}}\frac{\beta_s g_V}{2\sqrt{2}m_{\Sigma_c}}
\bar u(k_4,\lambda_4) u(k_2,\lambda_2) (k_2+k_4)^{\mu}\frac{-g_{\mu\nu}}{q_t^2-m_V^2}(k_1+k_3)^{\nu} \nonumber \\
& +\frac{\beta g_V}{\sqrt{2}}\frac{\lambda_S g_V}{3\sqrt{2}}
\bar u(k_4,\lambda_4) \gamma_{\mu}\gamma_{\nu} u(k_2,\lambda_2)
\Big( q_t^{\mu}\frac{-g^{\nu\sigma}}{q_t^2-m_V^2}(k_1+k_3)_{\sigma} - q_t^{\nu}\frac{-g^{\mu\sigma}}{q_t^2-m_V^2}(k_1+k_3)_{\sigma}        \Big), \\
V_S=&
2g_{S} m_D \ell_S \bar u(k_4,\lambda_4)\frac{1}{q_t^2-m_S^2} u(k_2,\lambda_2).
\end{align}

$\Sigma_c \bar{D}^* \to \Sigma_c \bar{D}^*$ interaction, 
\begin{align}
V_P=&
\frac{g}{f_{\pi}}\frac{g_1}{4m_{\Sigma_c}f_{\pi}}\epsilon^{\mu\nu\alpha\beta}\bar u(k_4,\lambda_4) (k_2+k_4)_{\beta} \gamma_{\mu}\gamma_{\alpha} q_{t\nu} u(k_2,\lambda_2) \frac{1}{q_t^2-m_P^2} \epsilon^{\rho\sigma\theta\tau}(k_1+k_3)_{\rho}  \epsilon_{\sigma}^*(k_3,\lambda_3) \epsilon_{\tau}(k_1,\lambda_1)q_{t\theta},\\
\nonumber V_V=&\frac{\beta g_V}{\sqrt{2}}\frac{\beta_s g_V}{2\sqrt{2}m_{\Sigma_c}}
\bar u(k_4,\lambda_4) u(k_2,\lambda_2) (k_2+k_4)^{\mu}\frac{-g_{\mu\nu}}{q_t^2-m_V^2}(k_1+k_3)^{\nu} \epsilon_{\alpha}^*(k_3,\lambda_3) \epsilon^{\alpha}(k_1,\lambda_1)\\
&\nonumber - \lambda g_Vm_{D^*}\frac{\beta_sg_V}{m_{\Sigma_c}} \bar u(k_4,\lambda_4) u(k_2,\lambda_2) (k_2+k_4)^{\mu}\frac{-g_{\mu\nu}}{q_t^2-m_V^2}\Big(q_{t\alpha}\epsilon^{*\alpha}(k_3,\lambda_3) \epsilon^{\nu}(k_1,\lambda_1) -  q_{t\alpha}\epsilon^{*\nu}(k_3,\lambda_3) \epsilon^{\alpha}(k_1,\lambda_1)      \Big) \\
\nonumber&-\frac{\beta g_V}{\sqrt{2}}\frac{\lambda_S g_V}{3\sqrt{2}}
\bar u(k_4,\lambda_4)\gamma_{\mu}\gamma_{\nu} u(k_2,\lambda_2)\Big(q_t^{\mu}\frac{-g^{\nu\alpha}}{q_t^2-m_V^2}-q_t^{\nu}\frac{-g^{\mu\alpha}}{q_t^2-m_V^2}\Big)(k_1+k_3)_{\alpha} \epsilon_{\beta}^*(k_3,\lambda_3) \epsilon^{\beta}(k_1,\lambda_1) \\
&\nonumber+(2\sqrt{2}\lambda g_Vm_{D^*})\frac{\lambda_S g_V}{3\sqrt{2}}\bar u(k_4,\lambda_4)\gamma_{\mu}\gamma_{\nu} u(k_2,\lambda_2) \Big(q_t^{\mu}\frac{-g^{\nu\alpha}}{q_t^2-m_V^2}-q_t^{\nu}\frac{-g^{\mu\alpha}}{q_t^2-m_V^2}\Big) \\ &\times \Big(q_t^{\beta}\epsilon_{\beta}^*(k_3,\lambda_3) \epsilon_{\alpha}(k_1,\lambda_1)-q_t^{\beta}\epsilon_{\alpha}^*(k_3,\lambda_3) \epsilon_{\beta}(k_1,\lambda_1)\Big),\\
V_S=&
-2g_{S} m_{D^*} \ell_S  \bar u(k_4,\lambda_4) \frac{1}{q_t^2-m_S^2} u(k_2,\lambda_2)
\epsilon_{\mu}^*(k_3,\lambda_3) \epsilon^{\mu}(k_1,\lambda_1).
\end{align}

$\Lambda_c \bar{D} \to \Lambda_c \bar{D}$ interaction, 
\begin{align}
V_V=&\frac{\beta g_V}{\sqrt{2}}\frac{\beta_B g_V}{2\sqrt{2}m_{\Lambda_c}}
\bar u(k_4,\lambda_4) u(k_2,\lambda_2) (k_2+k_4)^{\mu}\frac{-g_{\mu\nu}}{q_t^2-m_V^2}(k_1+k_3)^{\nu},\\
V_S=&
-2g_{S} m_D\ell_B \bar u(k_4,\lambda_4) \frac{1}{q_t^2-m_S^2} u(k_2,\lambda_2).
\end{align} 

$\Lambda_c \bar{D}^* \to \Lambda_c \bar{D}^*$ interaction, 
\begin{align}
\nonumber V_V=&-\frac{\beta g_V}{\sqrt{2}}\frac{\beta_B g_V}{2\sqrt{2}m_{\Lambda_c}}
\bar u(k_4,\lambda_4) u(k_2,\lambda_2) (k_2+k_4)^{\mu}\frac{-g_{\mu\nu}}{q_t^2-m_V^2}(k_1+k_3)^{\nu} \epsilon_{\alpha}^*(k_3,\lambda_3) \epsilon^{\alpha}(k_1,\lambda_1)
+ \lambda g_Vm_{D^*}\frac{\beta_Bg_V}{m_{\Lambda_c}} \\
&\times \bar u(k_4,\lambda_4) u(k_2,\lambda_2) (k_2+k_4)^{\mu}\frac{-g_{\mu\nu}}{q_t^2-m_V^2}\Big(q_{t\alpha}\epsilon^{*\alpha}(k_3,\lambda_3) \epsilon^{\nu}(k_1,\lambda_1) -  q_{t\alpha}\epsilon^{*\nu}(k_3,\lambda_3) \epsilon^{\alpha}(k_1,\lambda_1)      \Big),\\
V_S=&
2g_{S} m_{D^*}\ell_B  \bar u(k_4,\lambda_4) \frac{1}{q_t^2-m_S^2} u(k_2,\lambda_2)
\epsilon_{\mu}^*(k_3,\lambda_3) \epsilon^{\mu}(k_1,\lambda_1).
\end{align} 

$\Sigma_c \bar{D} \to \Sigma_c \bar{D}^*$ interaction,
\begin{align}
V_P=&
\frac{i2g\sqrt{m_Dm_{D^*}}}{f_{\pi}}\frac{g_1}{4m_{\Sigma_c}f_{\pi}}\epsilon^{\mu\nu\alpha\beta}(k_2+k_4)_{\beta} q_{t\nu} \bar u(k_4,\lambda_4)\gamma_{\mu}\gamma_{\alpha} u(k_2,\lambda_2)\frac{1}{q_t^2-m_P^2} 
\epsilon_{\sigma}^*(k_3,\lambda_3)q_t^{\sigma}, \\
\nonumber V_V=&-i\sqrt{2}\lambda g_V\frac{\beta_s g_V}{2\sqrt{2}m_{\Sigma_c}}
\bar u(k_4,\lambda_4) u(k_2,\lambda_2) (k_2+k_4)_{\mu}\frac{-g^{\mu\nu}}{q_t^2-m_V^2}\epsilon_{\lambda\alpha\nu\sigma} 
\epsilon^{*\sigma}(k_3,\lambda_3) (k_1+k_3)^{\lambda}q_t^{\alpha}\\
&+i\sqrt{2}\lambda g_V\frac{\lambda_S g_V}{3\sqrt{2}}\bar u(k_4,\lambda_4)\gamma_{\mu}\gamma_{\nu} u(k_2,\lambda_2) \Big(q_t^{\mu}\frac{-g^{\nu\sigma}}{q_t^2-m_V^2}-q_t^{\nu}\frac{-g^{\mu\sigma}}{q_t^2-m_V^2}\Big) 
\epsilon_{\lambda\alpha\sigma\rho} 
\epsilon^{*\rho}(k_3,\lambda_3) (k_1+k_3)^{\lambda}q_t^{\alpha}. 
\end{align}

$\Lambda_c \bar{D} \to \Sigma_c \bar{D}$ interaction,
\begin{align}
V_V=-i\frac{\beta g_V}{\sqrt{2}}\frac{\lambda_I g_V}{2\sqrt{6}\sqrt{m_{\Sigma_c}m_{\Lambda_c}}} (k_1+k_3)^{\mu}\Big( q_{t\lambda}\frac{-g_{\mu\kappa}}{q_t^2-m_V^2} - q_{t\kappa}\frac{-g_{\mu\lambda}}{q_t^2-m_V^2}   \Big)
\epsilon^{\alpha\beta\lambda\kappa}(k_2+k_4)_{\alpha} \bar u(k_4,\lambda_4) \gamma^5 \gamma_{\beta} u(k_2,\lambda_2).   
\end{align}

$\Lambda_c \bar{D}^* \to \Sigma_c \bar{D}$ interaction,
\begin{align}
V_P=&-\frac{2g\sqrt{m_Dm_{D^*}}}{f_{\pi}}\frac{g_4}{\sqrt{3}f_{\pi}}  \epsilon^{\mu}(k_1,\lambda_1)q_{t\mu}\frac{1}{q_t^2-m_P^2}
\bar u(k_4,\lambda_4) \gamma^5( \gamma^{\beta}+v^{\beta})q_{t\beta}u(k_2,\lambda_2),\\
 V_V=&\sqrt{2}\lambda g_V\frac{\lambda_I g_V}{2\sqrt{6}\sqrt{m_{\Sigma_c}m_{\Lambda_c}}} \epsilon_{\lambda\alpha\beta\mu} \epsilon^{\mu}(k_1,\lambda_1)(k_1+k_3)^{\lambda}q_t^{\alpha}\Big(\frac{-g^{\beta\kappa}}{q_t^2-m_V^2} q_t^{\rho} -\frac{-g^{\beta\rho}}{q_t^2-m_V^2} q_t^{\kappa}  \Big) \nonumber \\ &\times (k_2+k_4)^{\nu} \epsilon_{\nu\sigma\rho\kappa} \bar u(k_4,\lambda_4) \gamma^5 \gamma^{\sigma} u(k_2,\lambda_2).
\end{align}

$\Lambda_c \bar{D} \to \Sigma_c \bar{D}^*$ interaction,
\begin{align}
V_P=&-\frac{2g\sqrt{m_Dm_{D^*}}}{f_{\pi}}\frac{g_4}{\sqrt{3}f_{\pi}}  \epsilon^{*\mu}(k_3,\lambda_3)q_{t\mu}\frac{1}{q_t^2-m_P^2} \bar u(k_4,\lambda_4) \gamma^5( \gamma^{\beta}+v^{\beta})q_{t\beta}u(k_2,\lambda_2), \\
 V_V=&\sqrt{2}\lambda g_V\frac{\lambda_I g_V}{2\sqrt{6}\sqrt{m_{\Sigma_c}m_{\Lambda_c}}} \epsilon_{\lambda\alpha\beta\mu} \epsilon^{*\mu}(k_3,\lambda_3)(k_1+k_3)^{\lambda}q_t^{\alpha}\Big(\frac{-g^{\beta\kappa}}{q_t^2-m_V^2} q_t^{\rho} -\frac{-g^{\beta\rho}}{q_t^2-m_V^2} q_t^{\kappa}  \Big) \nonumber \\ &\times (k_2+k_4)^{\nu} \epsilon_{\nu\sigma\rho\kappa} \bar u(k_4,\lambda_4) \gamma^5 \gamma^{\sigma} u(k_2,\lambda_2).
\end{align}

$\Lambda_c \bar{D}^* \to \Sigma_c \bar{D}^*$ interaction,
\begin{align}
V_P=&i\frac{g}{f_{\pi}}\frac{g_4}{\sqrt{3}f_{\pi}} \epsilon_{\alpha\mu\nu\lambda} (k_1+k_3)^{\alpha}
\epsilon^{*\mu}(k_3,\lambda_3) \epsilon^{\lambda}(k_1,\lambda_1) q_t^{\nu}\frac{1}{q_t^2-m_P^2}
\bar u(k_4,\lambda_4) \gamma^5( \gamma^{\beta}+v^{\beta})q_{t\beta}u(k_2,\lambda_2), \\
\nonumber V_V=&\frac{i\beta g_V}{\sqrt{2}}\frac{\lambda_I g_V}{2\sqrt{6} \sqrt{m_{\Sigma_c}m_{\Lambda_c}}} 
\epsilon_{\alpha}^{*}(k_3,\lambda_3)\epsilon^{\alpha}(k_1,\lambda_1)
(k_1+k_3)_{\beta}(k_2+k_4)^{\mu} \Big(\frac{-g^{\beta\kappa}}{q_t^2-m_V^2} q_t^{\lambda} -\frac{-g^{\beta\lambda}}{q_t^2-m_V^2} q_t^{\kappa}  \Big) \epsilon_{\mu\nu\lambda\kappa} \bar u(k_4,\lambda_4)  \\ &\nonumber \times \gamma^5 \gamma^{\nu} u(k_2,\lambda_2) +i2\sqrt{2}\lambda g_Vm_{D^*} \frac{\lambda_I g_V}{2\sqrt{6}\sqrt{m_{\Sigma_c}m_{\Lambda_c}}} 
\epsilon^{*\sigma}(k_3,\lambda_3)\epsilon^{\rho}(k_1,\lambda_1) \Big(
q_{t\sigma}q_{t\lambda}\frac{-g_{\rho\kappa}}{q_t^2-m_V^2}-q_{t\sigma}q_{t\kappa}\frac{-g_{\rho\lambda}}{q_t^2-m_V^2}
\\
&-q_{t\rho}q_{t\lambda}\frac{-g_{\sigma\kappa}}{q_t^2-m_V^2}+q_{t\rho}q_{t\kappa}\frac{-g_{\sigma\lambda}}{q_t^2-m_V^2}
\Big) (k_2+k_4)_{\mu} \epsilon^{\mu\nu\lambda\kappa} \bar u(k_4,\lambda_4) \gamma^5 \gamma_{\nu} u(k_2,\lambda_2).
\end{align}

$\Lambda_c \bar{D} \to \Lambda_c \bar{D}^*$ interaction,
\begin{align}
V=i\sqrt{2}\lambda g_V\frac{\beta_B g_V}{2\sqrt{2}m_{\Lambda_c}}
\bar u(k_4,\lambda_4) u(k_2,\lambda_2) (k_2+k_4)_{\mu}\frac{-g^{\mu\nu}}{q_t^2-m_V^2}\epsilon_{\lambda\alpha\nu\sigma} 
\epsilon^{*\sigma}(k_3,\lambda_3) (k_1+k_3)^{\lambda}q_t^{\alpha}.
\end{align}

$\Sigma_c \bar{D} \to J/\psi p$ interaction, 
\begin{align}
V_D=&
\frac{ig_{DD\psi}g_{DN\Sigma_c}}{m_N+m_{\Sigma_c}}\bar u(k_4,\lambda_4) \gamma_5 q_t \!\!\!\!/ \,\, u(k_2,\lambda_2) \frac{1}{q_t^2-m_D^2}(k_1-q_t)^{\nu} \epsilon_{\nu}^*(k_3,\lambda_3),\\
V_{D^*}=&
{g_{DD^*\psi}}g_{D^*N\Sigma_c}\epsilon_{\alpha\beta\mu\nu}k_{3}^{\alpha}\epsilon^{*\beta}(k_3,\lambda_3)q_t^{\mu}\frac{-g^{\nu\sigma}}{q_t^2-m_{D^*}^2} \bar u(k_4,\lambda_4)\Big(\gamma_{\sigma}-\frac{i\kappa_{D^*N\Sigma_c}}{2m_N} \sigma_{\sigma\rho}q_{t}^{\rho}\Big)u(k_2,\lambda_2). 
\end{align}

$\Sigma_c \bar{D}^* \to J/\psi p$ interaction,
\begin{align}
V_D=&
\frac{ig_{DD^*\psi}g_{DN\Sigma_c}}{m_N+m_{\Sigma_c}}\epsilon_{\alpha\beta\mu\nu}k_{3}^{\alpha}\epsilon^{*\beta}(k_3,\lambda_3)k_1^{\mu} \epsilon^{\nu}(k_1,\lambda_1)
\frac{1}{q_t^2-m_{D}^2} \bar u(k_4,\lambda_4) \gamma_5 q_t \!\!\!\!/ \,\, u(k_2,\lambda_2),\\
\nonumber V_{D^*}=&
-g_{\psi D^{*}D^{*}}g_{D^*N\Sigma_c} \epsilon^{*\mu}(k_3,\lambda_3)\epsilon^{\nu}(k_1,\lambda_1) \Big( 2k_{1\mu}g_{\nu\alpha}-k_{1\alpha}g_{\mu\nu} + k_{3\nu}g_{\mu\alpha}    \Big) \frac{-g^{\alpha\beta}+q_t^{\alpha}q_t^{\beta}/m_{D^*}^2}{q_t^2-m_{D^*}^2} \\
&\bar u(k_4,\lambda_4)\Big( \gamma_{\beta} - \frac{i\kappa_{D^*N\Sigma_c}}{2m_N} \sigma_{\beta\rho}q_t^{\rho}\Big) u(k_2,\lambda_2).   
\end{align}

For $\Lambda_c \bar{D}^{(*)} \to J/\psi p$ transition, the $D$ and $D^*$ exchange potentials 
have the same structure as $\Sigma_c \bar{D}^{(*)} \to J/\psi p$ transition potentials.

\section{$\gamma p \to \Sigma_c\bar D$ transition potentials}
\label{sec:exppotenB}
For $\gamma p \to \Sigma_c \bar D^{}$ transition potentials, we consider $t$-channel $\bar{D}$ and $\bar{D}^{*}$ exchanges, $s$-channel nucleon exchange, $u$-channel $\Sigma_c$ exchange and a contact term.
The transition potentials are given as
\begin{align}
V_{\bar{D}*}=&g_{\gamma DD^*} g_{D^*N\Sigma_c}\epsilon_{\mu\nu\alpha\beta}k_1^{\mu}\epsilon^{\nu}(k_1,\lambda_1)q_t^{\alpha}
\frac{-g^{\beta\rho}+q^{\beta}q^{\rho}/m_{D^*}^2}{q_t^2-m_{D*}^2} \bar{u}(k_4,\lambda_4) \Big( \gamma_{\rho}-\frac{i\kappa_{D^*N\Sigma_c}}{2m_N}\sigma_{\rho\theta} q_t^{\theta}  \Big){u}(k_2,\lambda_2),\\ 
V_{\bar{D}}=& \frac{-ie_{\bar{D}}g_{DN\Sigma_c}}{m_N+m_{\Sigma_c}} 
\bar{u}(k_4,\lambda_4) q_t\!\!\!\!/\, \gamma_5 {u}(k_2,\lambda_2) \frac{1}{q_t^2-m_{D}^2} (2k_3-k_1)_{\mu}\epsilon^{\mu}(k_1,\lambda_1), \\
V_{p}=&\frac{-ig_{DN\Sigma_c}}{m_N+m_{\Sigma_c}} 
\bar{u}(k_4,\lambda_4) k_3\!\!\!\!\!/\,\, \gamma_5 \frac{q_s\!\!\!\!\!/+m_N}{q_s^2-m_{N}^2} \Big( e_p\gamma_{\mu}+\frac{ie\kappa_p}{2m_N}\sigma_{\mu\nu}k_1^{\nu}
 \Big)\epsilon^{\mu}(k_1,\lambda_1){u}(k_2,\lambda_2) , \\
V_{\Sigma_c}=&\frac{-ig_{DN\Sigma_c}}{m_N+m_{\Sigma_c}} \bar{u}(k_4,\lambda_4) \Big( e_{\Sigma_c}\gamma_{\mu}\ +\frac{ie\kappa_{\Sigma_c}}{2m_{\Sigma_c}} \sigma_{\mu\nu}k_1^{\nu}\Big) \epsilon^{\mu}(k_1,\lambda_1) \frac{q_u\!\!\!\!\!/+m_{\Sigma_c}}{q_u^2-m_{\Sigma_c}^2}
k_3\!\!\!\!\!/\,\,\gamma_5 {u}(k_2,\lambda_2), \\
V_C=&\frac{ie_{\bar{D}}g_{DN\Sigma_c}}{m_N+m_{\Sigma_c}} 
\bar{u}(k_4,\lambda_4) \gamma_{\mu} \gamma_5 {u}(k_2,\lambda_2)\epsilon^{\mu}(k_1,\lambda_1).
\end{align} 
The potential $V_{\bar{D}^*}$ is gauge invariant individually. With the replacement $\epsilon^{\mu}(k_1,\lambda_1) \to k_1^{\mu}$ in the potentials, the sum of all contributions $\widetilde{V}_{\bar{D}}$,
$\widetilde{V}_{p}$, $\widetilde{V}_{\Sigma_c}$ and $\widetilde{V}_{C}$
is 
\begin{align}
\widetilde{V}_{\bar{D}}+\widetilde{V}_{p}+\widetilde{V}_{\Sigma_c}+\widetilde{V}_{C}\,\,\sim\,\, (e_p-e_{\Sigma_c}-e_{\bar{D}})\bar{u}(k_4,\lambda_4) k_3\!\!\!\!\!/\,\,\gamma_5 {u}(k_2,\lambda_2),
\end{align} 
this vanishes as long as charge is conserved.

\section{$\gamma p \to \Sigma_c \bar D^*$ transition potentials}
\label{sec:exppotenC}
For $\gamma p \to \Sigma_c\bar D^{*}$ transition potentials, the $t$-channel $\bar{D}$ and $\bar{D}^{*}$ exchanges, $s$-channel nucleon exchange, $u$-channel $\Sigma_c$ exchange and a contact term are considered.
The transition potentials are given as
\begin{align}
V_{\bar{D}}=&\frac{ig_{\gamma DD^*}g_{DN\Sigma_c}}{m_N+m_{\Sigma_c}}
\epsilon^{\mu\nu\alpha\beta}k_{1\mu}\epsilon_{\nu}(k_1,\lambda)k_{3\alpha}\epsilon_{\beta}^*(k_3,\lambda_3)
\frac{1}{q_t^2-m_D^2}\bar{u}(k_4,\lambda_4)q_t\!\!\!\!/\, \gamma_5  {u}(k_2,\lambda_2),\\ 
\nonumber V_{\bar{D}^*} =& \frac{e_{\bar{D}^*}^{} g_{D^* N\Sigma_c}
}{q_t^2-m_{D^{*}}^2} \varepsilon_\nu^*(k_3,\lambda_3) 
\Big( 2k_3^\mu g^{\nu\alpha} - k_3^\alpha g^{\mu\nu} + k_1^\nu
  g^{\mu\alpha} \Big) \bar{u}(k_4,\lambda_4)
  \Big( g_{\alpha\beta} - \frac{q_{t\alpha}
    q_{t\beta}}{m_{D^*}^2}\Big)
\Big( \gamma^\beta -
  \frac{i\kappa_{D^*N\Sigma_c} }{2m_N} \sigma^{\beta \delta}
  q_{t\delta} \Big) \nonumber \\
  &\times {u}(k_2,\lambda_2)\epsilon_{\mu}(k_1,\lambda_1),
  \\ 
V_{\Sigma_c} =&g_{D^* N\Sigma_c}
\bar{u}(k_4,\lambda_4) \Big(e_{\Sigma_c}\gamma^\mu+
\frac{ie\kappa_{\Sigma_c} }{2m_{\Sigma_c}} \sigma^{\mu\nu}
  k_{1\nu} \Big) \epsilon_{\mu}(k_1,\lambda_1)\frac{q_u\!\!\!\!\!/\,+m_{\Sigma_c}}{q_u^2-m_{\Sigma_c}^2}
\Big( \gamma^\alpha -
  \frac{i\kappa_{D^*N\Sigma_c} }{2m_N} \sigma^{\alpha \beta}
  k_{3\beta} \Big) {u}(k_2,\lambda_2)\epsilon_{\alpha}^*(k_3,\lambda_3), \\
V_p =& {g_{D^* N\Sigma_c} }
\epsilon_{\mu}^*(k_3,\lambda_3) \bar{u}(k_4,\lambda_4)     
\Big( \gamma^{\mu} - \frac{i\kappa_{D^*N\Sigma_c} }{2m_N}
{\sigma^{\mu \alpha}}k_{3\alpha} \Big )
\frac{(q_s\!\!\!\!\!/\,+m_N)}{q_s^2-m_N^2}                         
\Big( e_p \gamma^\nu +\frac{ie \kappa_p}{2m_N}
\sigma^{\nu\beta} k_{1\beta} \Big) {u}(k_2,\lambda_2) \epsilon_{\nu}(k_1,\lambda_1) , \\
V_{C} =& -\frac{ie_{\bar{D}^*}g_{D^* N\Sigma_c} \kappa_{D^*N\Sigma_c}}{2m_N}
\epsilon_{\nu}^*(k_3,\lambda_3) \bar{u}(k_4,\lambda_4)     
\sigma^{\mu\nu} {u}(k_2,\lambda_2) \epsilon_{\mu}(k_1,\lambda_1) , 
\end{align}
The potential $V_{\bar{D}}$ is gauge invariant individually. With the replacement $\epsilon^{\mu}(k_1,\lambda_1) \to k_1^{\mu}$ in the potentials, the sum of all contributions $\widetilde{V}_{\bar{D}^*}$,
$\widetilde{V}_{p}$, $\widetilde{V}_{\Sigma_c}$ and $\widetilde{V}_{C}$
is 
\begin{align}
\widetilde{V}_{\bar{D}^*}+\widetilde{V}_{p}+\widetilde{V}_{\Sigma_c}+\widetilde{V}_{C}\,\,\sim \,\, (e_p-e_{\Sigma_c}-e_{\bar{D}^{*}}) \bar{u}(k_4,\lambda_4)     
\Big( \gamma^{\mu} - \frac{i\kappa_{D^*N\Sigma_c} }{2m_N}
{\sigma^{\mu \alpha}}k_{3\alpha} \Big ) {u}(k_2,\lambda_2),
\end{align}
this vanishes as long as charge is conserved.

For $\gamma p \to\Lambda_c \bar{D}^{(*)} $ transition, the potentials 
have the same structure as $\gamma p \to\Sigma_c \bar{D}^{(*)}$ transition potentials.

\end{widetext}

\bibliography{ref.bib}
\end{document}